\documentclass[11pt,twoside]{atmp}

\usepackage{amsmath,epsf,amssymb,latexsym,amsthm,setspace,bbm,array}

\usepackage[matrix,arrow,frame,import,curve,color]{xy}

\DeclareFontFamily{U}{rsf}{}
\DeclareFontShape{U}{rsf}{m}{n}{
  <5> <6> rsfs5 <7> <8> <9> rsfs7 <10-> rsfs10}{}
\DeclareMathAlphabet\Scr{U}{rsf}{m}{n}

\def\C{{\mathbb C}}

\def\P{{\mathbb P}}
\def\R{{\mathbb R}}
\def\Z{{\mathbb Z}}

\def\End{\operatorname{End}}

\def\SO{\operatorname{SO}}

\def\GU{\operatorname{U{}}}

\def\GE{\operatorname{E}}

\def\p{\partial}

\def\la{\langle}
\def\ra{\rangle}

\def\ff#1#2{{\textstyle\frac{#1}{#2}}}
\def\half{\frac{1}{2}}

\def\cA{{\mathcal A}}
\def\cB{{\mathcal B}}

\def\cD{{\mathcal D}}
\def\cE{{\mathcal E}}
\def\cF{{\mathcal F}}
\def\cG{{\mathcal G}}
\def\cH{{\mathcal H}}

\def\cJ{{\mathcal J}}
\def\cK{{\mathcal K}}
\def\cL{{\mathcal L}}

\def\cO{{\mathcal O}}

\def\cQ{{\mathcal Q}}

\def\cS{{\mathcal S}}

\def\cW{{\mathcal W}}

\def\ep{{\epsilon}}

\newcommand\alphab{\overline{\alpha}}

\newcommand\gammab{\overline{\gamma}}

\newcommand\etab{\overline{\eta}}
\newcommand\thetab{\overline{\theta}}

\newcommand\lambdab{\overline{\lambda}}

\newcommand\phib{\overline{\phi}}
\newcommand\chib{\overline{\chi}}
\newcommand\psib{\overline{\psi}}

\newcommand\ept{\widetilde{\ep}}

\newcommand\nut{\widetilde{\nu}}

\newcommand\phit{\widetilde{\phi}}

\newcommand\psit{\widetilde{\psi}}

\newcommand\vphi{\varphi}

\newcommand\Sigmab{\overline{\Sigma}}

\newcommand\Gammat{\widetilde{\Gamma}}

\newcommand\gh{\widehat{g}}

\newcommand\ab{\overline{a}}
\newcommand\bb{\overline{b}}
\newcommand\cb{\overline{c}}

\newcommand\qb{\overline{q}}

\newcommand\zb{\overline{z}}

\newcommand\bt{\widetilde{b}}

\newcommand\htld{\widetilde{h}} 

\newcommand\Gh{\widehat{G}}

\newcommand\Kh{\widehat{K}}

\newcommand\Mh{\widehat{M}}

\newcommand\Oh{\widehat{O}}

\newcommand\Pb{\overline{P}}

\newcommand\Sb{\overline{S}}

\newcommand\Zb{\overline{Z}}

\newcommand\Pt{\widetilde{P}}

\def\GUL{\GU(1)_{\text{L}}}
\def\GUR{\GU(1)_{\text{R}}}
\def\GUG{\GU(1)_{\text{G}}}

\def\CO#1#2{{[#1,#2]}}

\def\rep#1{{{\boldsymbol{#1}}}}
\def\brep#1{{{\overline{\boldsymbol{#1}}}}}

\def\bt{{{\boldsymbol{t}}}}
\def\bzer{{{\boldsymbol{0}}}}

\def\cDb{{\overline{\cD}}}
\def\cQb{{\overline{\cQ}}}
\def\cGb{{\overline{\cG}}}

\def\cAh{{\widehat{\cA}}}

\def\mon{{{\mathsf{M}}}}
\def\monb{{{\overline{\mon}}}}

\def\Sym{\operatorname{Sym}}

\begin{document}

\title[Old issues and linear sigma models]
{Old issues and linear sigma models}


\author[McOrist and Melnikov]{Jock McOrist$^a$ and Ilarion V. Melnikov$^b$}
\address{ $^a$ DAMTP, Centre for Mathematical Sciences \\
 University of Cambridge \\
 Wilberforce Rd, Cambridge, CB3 0WA, UK}
 \addressemail{j.mcorist@damtp.cam.ac.uk}
 
 \address{ $^b$ Max-Planck-Institut f\"ur Gravitationsphysik (Albert-Einstein-Institut)\\
 Am M\"uhlenberg 1, D-14476 Golm, Germany}  
\addressemail{ilarion@aei.mpg.de}

\begin{abstract}
Using mirror symmetry, we resolve an old puzzle in the linear sigma model description of
the spacetime Higgs mechanism in a heterotic string compactification with (2,2) worldsheet supersymmetry.  The resolution has a nice spacetime interpretation via the normalization of physical fields and suggests that with a little care deformations of the linear sigma model can describe heterotic Higgs branches.
\end{abstract}

\maketitle
\cutpage 
\setcounter{page}{2}

\noindent

\section{Introduction}
Calabi-Yau compactifications of the perturbative heterotic string
to $d=4$ Minkowski space with $N=1$ spacetime supersymmetry
occupy a prominent position in the space of string vacua.  To the
chagrin of the phenomenologist these models have a large
number of moduli, which precludes direct applications to the real world;
however, this very same feature means that many properties of these
models are readily computable and can give new insights into
general features of heterotic compactifications away from the supergravity limit.

The gauged linear sigma model (GLSM)~\cite{Witten:1993yc} has proven to be an important tool in the exploration of heterotic moduli spaces. It is particularly important in the studies of vacua admitting a large radius description as stable holomorphic vector bundles over Calabi-Yau complete intersections in  toric varieties. The main utility of the GLSM is the
presentation of at least some of the exactly marginal deformations of the (0,2) worldsheet superconformal field theory (SCFT)  as parameters in
a weakly coupled UV Lagrangian. This presentation, when combined with quasi-topological field theory techniques, can be used to argue that certain deformations are exactly marginal, to compute physically interesting correlators, and to connect different regions in the moduli space.  A recent review of this approach may be found in~\cite{McOrist:2010ae}.

Typically, the GLSM studies carried out to date have focused on deformations that preserve the rank of the holomorphic bundle. In spacetime this is tantamount to ignoring deformations along the Higgs branch.  The aim of this note is to explain that the GLSM can also be used to probe the Higgs branch, at least in the simplest situation, where the undeformed theory is a compactification of the $\GE_8\times\GE_8$ heterotic string with (2,2) worldsheet supersymmetry. In that case, in the large radius limit, the stable bundle is just the tangent bundle of the Calabi-Yau manifold.  For generic values of the (2,2) moduli the gauge group is $\GE_6\times\GE_8$, and we seek to describe the Higgs mechanism for the $\GE_6$ factor.

The main result of this work is a resolution of a puzzle, first raised in~\cite{Distler:1993mk}, concerning rank-changing deformations of the most venerable model of all --- the quintic hypersurface in $\P^4$.  In brief, the issue is this:  it is well established that this compactification has Higgs branch deformations breaking $\GE_6 \to \SO(10)$~\cite{Witten:1985bz,Dine:1988kq}; there is an obvious guess as to how these deformations are incorporated in the GLSM~\cite{Distler:1993mk,Distler:1995mi}; yet the application of standard GLSM tools yields an inconsistent massless spectrum at the Landau-Ginzburg locus of the deformed model!

This is confusing to say the least, and it might dampen one's
enthusiasm for applying GLSM tools to explore Higgs branch
deformations.  Fortunately, there is a simple resolution.  We
will describe how to deform the GLSM to obtain the desired
deformations and, by using mirror symmetry, check that the
puzzle is resolved in the full GLSM.  The mirror perspective will
also identify the basic problem:  a subtlety in the reduction of the
GLSM to the simpler Landau-Ginzburg description.

While we show that care is required in using the GLSM to describe the
Higgs branch, there are arguments that remain unmodified by turning
on the deformation.  For example the results of~\cite{Beasley:2003fx}
imply that the Higgs deformations are not lifted
by worldsheet instanton effects.  With a little bit
of care the GLSM can be used to study the Higgs branch
and continues to be a powerful and versatile tool.

The rest of the note is summarized as follows.  In section~\ref{s:genquin}
we review some standard facts about the (2,2) SCFT defined by the
quintic hypersurface and its space of deformations, and we summarize some
more general results.  Next, in section~\ref{s:glsmdef},
we turn to the quintic GLSM;  we identify a natural set of infinitesimal Higgs
deformations and discuss some
generalizations of the construction.  In section~\ref{s:LG} we tackle the
Landau-Ginzburg puzzle, and we end with a brief outlook.

\section{Quintic lore and its generalizations}
\label{s:genquin}
As originally introduced in~\cite{Candelas:1985en}, the quintic compactification preserves (2,2) supersymmetry on the worldsheet.  The (2,2)
SCFT has a special K\"ahler moduli space, which locally
splits into a product of the complexified K\"ahler and complex
structure moduli spaces of dimensions $1$ and $101$ respectively.  The Fermat quintic, defined by the vanishing of
$\sum_i Z_i^5 = 0$ in $\P^4$, exhibits a global symmetry,
$G = (S_5 \ltimes \Z_5^5)/\Z_5$.  The $S_5$ is generated by permuting the $\P^4$ coordinates, while the $\Z_5^5$ maps $Z_i \mapsto e^{2\pi i a_i/5} Z_i$, with $a_i  = 0,\ldots,4$; since a diagonal phase rotation leaves $\P^4$ invariant, it does not lead to a symmetry.  This large symmetry group exists for all values of the K\"ahler modulus,
connecting the large radius limit and the Gepner point~\cite{Gepner:1987vz}.  The charged matter is organized in $\rep{27}$ and $\brep{27}^{\oplus 101}$
representations of $\GE_6$,  whose vertex operators are in one-to-one correspondence with
the (2,2) moduli and thus remain massless at every smooth point in the
(2,2) moduli space~\cite{Dixon:1987bg}.  In addition to these massless fields, analysis in the large radius limit identifies $224$ massless $\GE_6$ singlets.  Since these have a geometric interpretation as elements of $H^1(\End T)$, i.e. the infinitesimal deformations of the Calabi-Yau tangent
bundle, we will refer to them as ``bundle singlets.''  These singlets can also be identified at the Gepner point,
leading to one of the earliest indications that (0,2) compactifications can possess remarkable (from the low energy point of view) non-renormalization properties.

A natural question to ask~\cite{Witten:1985bz} is whether the theory has  flat
directions along which the (2,2) worldsheet supersymmetry is broken to (0,2) --- the minimum necessary for an $N=1$ spacetime supersymmetric heterotic vacuum~\cite{Banks:1987cy}.  There are
two types of deformations to consider: one might either try to give a vacuum expectation value (VEV) to a bundle singlet or move onto the Higgs branch by giving VEVs to the $\rep{27}$ and $\brep{27}$s.  In the large radius limit the latter corresponds to deformations of $T\oplus \cO^{\oplus k}$, where $\cO$ is the trivial bundle over the Calabi-Yau.

The existence of a Higgs branch, at least for special values of complex structure moduli, is
established by a beautiful argument combining worldsheet and spacetime ideas~\cite{Dine:1988kq}.  As it is rather important for our purposes to feel sure that this branch exists, we will review this argument.  From the spacetime point of view an infinitesimal Higgs deformation may be obstructed by D- or F-terms.  In the case of breaking $\GE_6\to \SO(10)$, we expect that the former can be made to vanish by
judiciously relating the $\rep{27}$ and $\brep{27}$ VEVs.  To examine the F-terms obstructions,  the authors of~\cite{Dine:1988kq} note\footnote{See
also~\cite{Dine:1986zy,Dine:1987vf,Dine:1987bq} for related discussions.} that when the unperturbed vacuum is defined by a CFT (as opposed to a sigma-model expansion around the infinite distance large radius limit), then the effective superpotential for the massless fields should be given by a power-series in the fields.  Thus, if the F-term obstruction vanishes to all orders in the fields, it must vanish exactly.

Suppose then that the complex structure moduli are tuned to the $G$-preserving locus, and we consider an infinitesimal deformation that gives VEVs to the $\rep{27}$ and $\brep{27}_0$ --- the multiplet corresponding to the unique permutation-invariant monomial $Z_1 Z_2 Z_3 Z_4 Z_5$.
The possible F-term obstructions are of the form
$(\rep{27}\cdot \brep{27}_0)^k$ and $S (\rep{27} \cdot \brep{27}_0)^k$, where $S$ is any singlet.
However, some of the G-transformations act as discrete R-symmetries of the unperturbed theory, and these
R-symmetries rule out both types of couplings.\footnote{These constraints are discussed in appendix~\ref{app:LG}, where we slightly generalize the result.}  Since the obstructions vanish to all orders, we conclude that the deformation can be integrated to a flat Higgs direction.

The discrete R-symmetries identified in~\cite{Dine:1988kq} also imply that the bundle singlets
remain massless on the $G$-preserving locus of the moduli space, thereby providing a low energy explanation for the seemingly miraculous absence of otherwise allowed F-terms.
Nevertheless, a more delicate reasoning~\cite{Distler:1994hq,Berglund:1995yu,Silverstein:1995re} demonstrates
that (0,2) models appear to be string miracles:  the bundle singlets have a flat potential for all values of the (2,2) moduli!
These results have since been understood in a more general context of the
GLSM:  since all of the $\GE_6$-neutral singlets are represented as deformations of
a GLSM Lagrangian, one can apply the arguments
of~\cite{Silverstein:1995re,Basu:2003bq,Beasley:2003fx} to show that every one of the
$326$ singlets constitutes a flat direction.

These arguments can be generalized to many more (2,2)
and (0,2) compactifications with a GLSM description.  A (2,2) model with a GLSM description naturally includes three types of $\GE_6$-preserving deformations:
the  (2,2)-preserving ``toric'' K\"ahler and the ``polynomial'' complex structure deformations,
as well as polynomial bundle deformations, which preserve (0,2) supersymmetry and in a large radius limit correspond to unobstructed deformations of the tangent bundle.  In general bundle deformations can be lifted by worldsheet
instantons~\cite{Dine:1986zy}.  However, in the GLSM context the
possible lifting is highly constrained by the results of~\cite{Silverstein:1995re,Berglund:1995yu,Basu:2003bq,Beasley:2003fx}:  it is expected that worldsheet instanton corrections due to the toric K\"ahler moduli do not lift the deformations that are representable in the GLSM.

The results of~\cite{Silverstein:1995re,Basu:2003bq,Beasley:2003fx}
can also be applied in the more general context of generic (0,2) theories with a
GLSM description.  In favorable cases, e.g. models based on a stable bundle over
a Calabi-Yau manifold without non-toric K\"ahler parameters, these arguments should
be sufficient to show that the compactification is not destabilized by worldsheet instantons.

Having assured ourselves that we stand on reasonably firm ground, we will now discuss how to construct Higgs deformations in the GLSM.  To that end, we will first discuss
the unperturbed theory.

\section{The  quintic GLSM}
\label{s:glsmdef}
We begin with the familiar structure of the GLSM for the (2,2) supersymmetric
compactification of the quintic.  It will be convenient to state the field content in
terms of (0,2) multiplets.  We have bosonic chiral multiplets $\Sigma$, $\Phi^0$, and  $\Phi^i$, $i=1,\ldots, 5$, as well as Fermi multiplets $\Gamma^0$ and $\Gamma^i$.  The latter are not
chiral but instead obey the constraints $\cDb_+ \Gamma = E(\Phi,\Sigma)$.  These fields
are coupled to a (0,2) vector multiplet with a chiral Fermi field-strength multiplet $\Upsilon$.\footnote{We
are following here the standard description as in~\cite{Witten:1993yc} in the conventions of~\cite{McOrist:2008ji}.}  The superspace expansions of these multiplets are as follows:
\begin{align}
\Upsilon & =  -2 (\lambda_{-} - i \theta^+(D -i f_{01}) -i \theta^+\thetab^+ \p_+ \lambda_{-} ), \nonumber\\
\Phi &= \phi + \sqrt{2} \theta^+ \psi_+ -i \theta^+\thetab^+ \nabla_+ \phi, \\
\Sigma & =  \sigma +\sqrt{2} \theta^+\lambda_{+} - i \theta^+\thetab^+ \p_+\sigma,\nonumber\\
\Gamma &= \gamma_- - \sqrt{2} \theta^+ G - i \theta^+\thetab^+ \nabla_+\gamma_- - \sqrt{2}\thetab^+ E(\Phi,\Sigma). \nonumber
\end{align}
This is given in $(-,+)$ signature with gauge-covariant derivatives $\nabla_\pm$ and superspace derivatives
\begin{equation}
\cD_+  = \p_{\theta^+} - i \thetab^+ \nabla_+,~~~ \cDb_+ = -\p_{\thetab^+} + i \theta^+ \nabla_+.
\end{equation}
$D$ is the top component of the vector field multiplet, $f_{01}$ is the gauge field-strength, and the $G$ are auxiliary fields.

The Lagrangian is constrained by the $\GUG$ gauge symmetry, as well as a  non-anomalous $\GUL\times\GUR$ symmetry. The charges of the multiplets are indicated in
table~\ref{table:M0charges}, which also includes an additional Fermi multiplet
$\Gamma^6$.
\begin{table}
\begin{center}
\begin{tabular}{|c|c|c|c|c|c|c|c||c|}
\hline
$~$ 		&$\theta^+$	&$\Phi^i$	&$\Phi^0$	&$\Gamma^i$	&$\Gamma^0$ &$\Upsilon$	&$\Sigma$  &$\Gamma^6$	 \\ \hline
$\GUL$ 	&$0$			&$0$		&$1$ 	&$-1$		&$0$  		&$0$			&$-1$		&$-1$   \\ \hline
$\GUR$ 	&$1$			&$0$		&$1$ 	&$0$			&$1$  		&$1$			&$1$			 &$0$    \\ \hline
$\GUG$ 		&$0$			&$1$		&$-5$ 	&$1$			&$-5$  		&$0$			&$0$			 &$0$  \\ \hline
\end{tabular}
\caption{Fields and charges for the (2,2) quintic; $\theta^+$ is the chiral (0,2) superspace coordinate.}
\label{table:M0charges}
\end{center}
\end{table}
The Lagrangian consists of canonical kinetic terms, potential terms due to the chirality constraints on the $\Gamma$,
and a (0,2) superpotential
\begin{align}
\cW_0 = \ff{1}{4} \tau \Upsilon +  \Phi^0 \Gamma^i J_i(\Phi) + \Gamma^0 P(\Phi),
\end{align}
where $J_i$ and $P$ are polynomials in the $\Phi^i$ of charges, respectively, $4$ and $5$, and $\tau = ir +\theta/2\pi$
 is a holomorphic coupling combining the F-I parameter $r$ and
the theta angle.
The theory enjoys (0,2) supersymmetry provided $\cW_0$ is chiral, which requires
\begin{align}
\label{eq:susyconst}
 E^i J_i + E^0 P = 0.
\end{align}
The (0,2) supersymmetry is enhanced to (2,2) when the $E$ and $J$ couplings take on special values
$E^0 = -5 \Phi^0 \Sigma$, $E^i = \Phi^i \Sigma$, and $J_i = P_{,i}$.

The geometric import of this construction is well-known~\cite{Witten:1993yc,Distler:1995mi}.
When the F-I parameter $r \gg 0$, the low energy theory is
described by a (0,2) NLSM with target-space the Calabi-Yau hypersurface $M = \{ P = 0 \} \subset \P^4$ and massless
left-moving fermions coupled to a bundle $\cE$ defined as the cohomology of the complex
\begin{align}
\xymatrix{0 \ar[r] &\cO   \ar[r]^-{E^i} & \cO_M(1)^{\oplus 5} \ar[r]^-{J_i} & \cO_M(5)\ar[r] & 0 .}
\end{align}
When the $E$ and $J$ couplings take their (2,2) values it is easy to see that  $\cE = T_M$, and
the NLSM enjoys (2,2) supersymmetry.
It is believed that the IR limit of this theory defines a (2,2) SCFT with central
charge $9$ and integral $\GUL\times\GUR$ charges.

It is a text-book fact that such a SCFT
can be used to construct an $N=1$ spacetime supersymmetric heterotic compactification with
gauge group $\GE_6\times \GE_8$~\cite{Green:1987sp,Polchinski:1998rr}.  In brief, we need
the following additional ingredients:
four free (0,1) supermultiplets representing the $\R^{1,3}$ directions, ten free left-moving fermions
$\xi^\alpha$, a level $1$ left-moving $\GE_8$ algebra, and
the $bc-\beta\gamma$ ghost system of the critical heterotic string.
Performing a requisite GSO projection, we obtain our compactification.  The $\SO(10)\times\GUL$ left-moving currents constitute the linearly realized part of the $\GE_6$ gauge symmetry, with remaining gauge bosons coming from the twisted sectors of the GSO projection.  In our conventions the $\SO(10)\times \GUL$ decompositions
of relevant $\GE_6$ representations are as follows:
\begin{align}
\rep{78} &= \rep{16}_{-3/2} \oplus \rep{45}_0\oplus \rep{1}_0 \oplus \brep{16}_{3/2}, \nonumber\\
\rep{27} &= \rep{10_{-1}} \oplus \rep{16}_{1/2}\oplus \rep{1}_2, \\
\brep{27} &= \rep{1}_{-2} \oplus \brep{16}_{-1/2} \oplus \rep{10}_1~\nonumber.
\end{align}

\subsection{Deformations of the quintic theory}
\label{ss:quindef}
One of the main uses of the GLSM is to provide a tractable description of a subspace of the
moduli space of the SCFT.  For instance, in the (2,2) quintic compactification
the GLSM parameter $\tau$ corresponds to the complexified K\"ahler parameter of the SCFT, while the holomorphic
couplings in the quintic polynomial $P$, when taken modulo holomorphic field redefinitions, describe
the $101$ complex structure deformations of the quintic.  By varying the $E$ and $J$ couplings while
preserving~(\ref{eq:susyconst}) it is also possible to describe the $224$ deformations of
the tangent bundle of the quintic.

It turns out that components of the Higgs branch can also be given a GLSM description.  This is particularly simple in the context of (2,2) compactifications due to the well-known relation between the vertex operators
for neutral moduli and the charged matter fields~\cite{Dixon:1987bg,Dixon:1989fj}.
Let $O_a$ and $\Oh_m$ denote elements of the (a,c) and (c,c) rings with
$\GUL\times\GUR$ charges $(-1,1)$ and $(1,1)$, respectively.  The moduli are constructed
by acting on these elements with the left- and right-moving supercharges, which we denote,
respectively, by $\cG,\cGb$ and
$\cQ,\cQb$:
\begin{align}
M_a = \cGb_{-1/2} \cQ_{-1/2} \cdot O_a, \qquad
\Mh_m = \cG_{-1/2} \cQ_{-1/2} \cdot \Oh_m.
\end{align}
Similarly, the vertex operators for the $\rep{10}_{-1} \subset \rep{27}$ and $\rep{10}_1 \subset \brep{27}$
are obtained by replacing the $\cG,\cGb$ action with a multiplication by the free fermions:
\begin{align}
O_a \to \xi {\cQ}_{-1/2} \cdot O_a, \qquad
\Oh_m \to \xi {\cQ}_{-1/2} \cdot \Oh_m.
\end{align}
Thus, in the SCFT we have a simple way to give VEVs to components in $\rep{10}_{\pm 1}$:  we should
perturb the theory by
\begin{align}
\Delta S = - \ep^a_\alpha \int d^2 z~ \xi^\alpha \cQ_{-1/2} \cdot O_a - \ept^m_\alpha  \int d^2 z~ \xi^\alpha \cQ_{-1/2} \Oh_m +\text{h.c.},
\end{align}
where $\alpha$ runs over the ten free left-moving fermions, and the $\ep^a_\alpha$ and $\ept^m_\alpha$ denote
the deformation parameters.  Of course it is not so easy to determine which, if any, of these parameters can
integrated up to exactly marginal deformations.

Let us narrow our sights further on Higgs deformations breaking $\GE_6\to \SO(10)$.  In order to build
the $\SO(10)$ current algebra in the deformed SCFT we will need to combine the currents of a linearly
realized $\SO(8)\times \GUL$ algebra with contributions from the twisted sectors of the GSO projection. Hence our deformations should leave eight free left-moving $\xi$, and they should couple
to the remaining two $\xi$s in such a way as to preserve a $\GUL$ symmetry.  This is easily
done by combining the two coupled $\xi$s into a Weyl fermion $\gamma^6$ and writing our coupling as
\begin{align}
\label{eq:M1def}
\Delta S = -\ep^a   \int d^2 z ~\gammab^6 \cQ_{-1/2} \cdot O_a - \ept^m \int d^2 z ~\gamma^6 \cQ_{-1/2} \Oh_m + \text{h.c}.~.
\end{align}
Note that $\Delta S$ breaks the $\GUL$ symmetry of the undeformed (2,2) theory, as well as the $\GU(1)$ symmetry of the free $\gamma^6$; however a linear combination of the two, under which $\gamma^6$ transforms with charge $-1$ is preserved.  In what follows, we will refer to this as ``the'' $\GUL$ symmetry.  We will denote the (2,2) left-moving R-symmetry by $\GUL'$.

In a geometric setting, where the SCFT is realized by a NLSM,
this infinitesimal deformation has a simple interpretation:  the infinitesimal deformations of $T_M \oplus \cO_M$
are described by
\begin{align}
H^1(T_M\oplus\cO_M) = H^1(\End T_M) \oplus H^1(T^\ast_M) \oplus H^1(T_M),
\end{align}
and the $\ep$ and $\ept$
label the elements of $H^1(T^\ast_M)$ and $H^1(T_M)$, respectively.

We can now apply this idea in the context of a (2,2) GLSM and $\GE_6\to \SO(10)$ deformations:  all we need are GLSM representatives of the $O_a$ and $\Oh_m$.
In the example of the quintic, for instance, the (a,c) chiral operator is represented by $\sigma$, while the
$101$ (c,c) operators are represented by gauge-invariant polynomials $\phi^0 f (\phi^1,\ldots,\phi^5)$.  The abstract
couplings in $\Delta S$ now take a concrete form.  The quintic GLSM is supplemented by an additional Fermi
multiplet $\Gamma^6$ with a chiral constraint
\begin{align}
\cDb_+ \Gamma^6 = \ep \Sigma,
\end{align}
and the (0,2) superpotential is modified to
\begin{align}
\cW_0 \to \cW_1 =  \ff{1}{4}\tau \Upsilon+\Phi^0 (\Gamma^i J_i + \Gamma^6 J_6 ) + \Gamma^0 P,
\end{align}
where $J_6 = \ept^m f_m(\Phi)$ is a quintic polynomial.
This GLSM, which we dub M1, will be (0,2) supersymmetric provided that the couplings obey the (0,2)
supersymmetry constraint, which for $E^i$ and $E^0$ at their (2,2) values implies
\begin{align}
\label{eq:susyconstM1}
\Phi^i J_i + \ep J_6 = 5 P.
\end{align}

When $r \gg 0$, we expect M1 to reduce to a (0,2) NLSM for a rank $4$ bundle $\cE_1$ encoded
by the cohomology of the complex
\begin{align}
\label{eq:M1bundle}
\xymatrix{0 \ar[r] &\cO_M   \ar[r]^-{\binom{E^i}{\ep}} & \cO_M(1)^{\oplus 5}\oplus \cO_M \ar[r]^-{(J_i, J_6)} & \cO_M(5)\ar[r] & 0 .}
\end{align}
The argument of~\cite{Beasley:2003fx} can be easily applied here to show that  worldsheet instantons cannot destabilize the solution.  Thus, provided $\cE_1$ is a stable bundle, we expect the M1 GLSM to flow to a deformed SCFT describing a heterotic vacuum with $\SO(10)$ gauge symmetry.\footnote{Recall that the usual spacetime non-renormalization arguments rule out $\alpha'$ perturbative corrections.}

\subsection{Stability of $\cE_1$ via the M2 GLSM}
A simple and instructive way to demonstrate stability of $\cE_1$ is to consider a related GLSM
description, which we dub the M2 model. Consider the most general M1 GLSM for $\ep \neq 0$.  That is,
$E^i = A^i_j \Phi^j \Sigma$, $E^0 = -b \Phi^0 \Sigma$, and $J_6$ is determined by~(\ref{eq:susyconstM1}) in terms of $E$, $J_i$ and $P$. Now consider the following redefinition of the Fermi
multiplets:
\begin{align}
\Gamma^i = \Gammat^i + \ep^{-1}\Phi^j A^i_j \Gammat^6, \qquad
\Gamma^0 = \Gammat^0 -\ep^{-1} b \Phi^0\Gammat^6, \qquad
\Gamma^6 = \Gammat^6.
\end{align}
With this redefinition the (0,2) superpotential takes a simpler
form
\begin{align}
\cW_1 \mapsto \cW_2 = \Phi^0 \Gammat^i J_i + \Gammat^0 P.
\end{align}
The redefinition also acts on the $E$-couplings: $E^0$ and $E^i$ are set to zero,
while $E^6 = \ep \Sigma$.  Thus, up to presumably irrelevant modifications of kinetic
terms for the Fermi multiplets, the M1 GLSM consists of a free massive multiplet
$(\Sigma,\Gammat^6)$ and the remaining degrees of freedom $\Phi,\Gammat$
coupled to the $\GUG$ gauge field.  The latter defines the M2 GLSM.
Up to presumably irrelevant terms the M2 and M1 models only differ by a decoupled massive multiplet, and we expect that they lead to the same IR dynamics.  Note that we could have also obtained M2 from M1 by taking $\ep \to \infty$, while at the same time scaling $\ept \to 0$.

The M2 GLSM is a simpler theory: there are fewer fields, no $E$-couplings, and therefore no need for a (0,2) supersymmetry constraint.  The left-moving fermions couple
to a familiar rank $4$ monad bundle $\cE_2$, defined as a kernel
\begin{align}
\label{eq:M2bundle}
\xymatrix{0 \ar[r] &\cE_2   \ar[r]  & \cO_M(1)^{\oplus 5}\ar[r]^-{J_i} & \cO_M(5)\ar[r] & 0 .}
\end{align}
This bundle splits if and only if the defining quintic polynomial $P$ is in the ideal $\la J_1,\ldots, J_5\ra$,  in which case $\cE_2 = \cF \oplus \cO$ with $\cF$ a deformation of $T_M$.  When in addition $J_i = P_{,i}$, we find $\cE_2 = T_M\oplus \cO_M$.
The same change of variables that showed M2 = M1 in the IR for $\ep \neq 0$ also makes it obvious that $\cE_2 = \cE_1$ as holomorphic bundles.  The stability of $\cE_2$ with generic $J_i$ has been proven many times, e.g. in~\cite{Distler:1987ee,Huybrechts:1995tb,Li:2004hx}, so it seems that we can expect the IR limits of both M1 and M2 GLSMs to define heterotic vacua with gauge group $\SO(10)$.

Although M1 and M2 should lead to identical IR physics for $\ep \neq 0$, the situation
is not so clear for $\ep = 0$.  For instance, we might wonder whether we can describe
the (2,2) locus in the context of the M2 model.  Since the field redefinition relating M2 and M1 is singular at $\ep = 0$, one might suspect that this is not so simple.
Indeed, we cannot expect to find (2,2) supersymmetry in the UV GLSM without
``integrating in'' the missing massive fields.  However, one might hope that
the IR NLSM derived from the M2 GLSM fares better.  After all, by
setting $J_i = P_{,i}$ we do
obtain $\cE_2 = T_M \oplus \cO_M$.  As we discuss in appendix~\ref{app:m1red},
this does not seem to be the case, and recovery of the (2,2) locus may only
be possible at the level of the SCFT.

\subsection{Infinitesimal deformations of the M1 GLSM}
We can generalize the construction of the previous section to include the full set
of $\SO(10)$-preserving deformations encoded by the $E^i$ and $J_i$ potentials.
Following~\cite{Kreuzer:2010ph} we can count the infinitesimal deformations obtained in
this fashion.

The M1 Lagrangian depends on $630$ complex parameters,
of which $126$ are eliminated by the
(0,2) supersymmetry constraint.  The GLSM deformation space is obtained as a quotient of this $504$-dimensional
space by the $80$-dimensional space of holomorphic field redefinitions.  As not all redefinitions act
properly on the parameter space,\footnote{For instance, the $\GUL\times \GUG$ transformations leave the parameters
invariant.} care must be taken to obtain the correct count.  We find that if  $J_6 \in \la J_1,J_2,\ldots,J_5\ra$
and $\ep = 0$, there are $428$ infinitesimal GLSM deformations; otherwise the number drops down to
$427$.\footnote{For $\ep \neq 0$, this counting is reproduced in the M2 model, where we find $427$ deformations.}

In spacetime these deformations should be interpreted as (at least a subset of)
$\SO(10)$ singlets that remain massless for all values of the GLSM parameters.  When $\ep = 0$ and $J_6$ is in the ideal $\la J_1,J_2,\ldots, J_5\ra$, M1 is
equivalent by a field redefinition to the quintic GLSM supplemented by a free left-moving Weyl fermion, and
thus we expect $326+1+101 =428$ massless $\SO(10)$ singlets; at a more generic point, where the gauge group is
broken to $\SO(10)$, we expect to lose one singlet due to the Higgs mechanism and possibly additional
ones due to F-term mass terms.  The GLSM counting suggests there is no additional F-term lifting of the singlets.

\subsection{Generalizations}
The construction of $\GE_6\to \SO(10)$ Higgsing via GLSM deformations is easily generalized to (2,2)
compactifications where $M$ is a Calabi-Yau hypersurface in a toric variety $\{\C^{k+4}~\backslash~ F \} / (\C^\ast)^k$.  In this case, the ``toric'' K\"ahler moduli are represented by the $\sigma_a$ --- the scalars in the $\GUG^k$ gauge multiplets, while the ``polynomial'' complex structure deformations are represented by gauge-invariant monomials in the $\phi^i$~\cite{Morrison:1994fr}.  Using these operators as the building blocks, we can deform the initial $\GE_6$ GLSM to an $\SO(10)$ GLSM with a correspondingly simple generalization of the deformed bundle in~(\ref{eq:M1bundle}).  Similarly, it should not be too difficult to generalize the construction to rank $5$ cases, as well as  Calabi-Yau complete intersections in toric varieties.  However, it
should be borne in mind that, as in the case of (2,2) deformations, the number of infinitesimal GLSM deformations may not accurately reflect the number of massless singlets --- the quintic example is particularly fortuitous in this sense.

Lacking a generalization of~\cite{Li:2004hx} to these more general hypersurfaces and complete intersections, one must provide a separate argument that the construction leads to a stable deformation of $T_M \oplus \cO_M^{\oplus k}$.  It would be nice to have a general geometric statement; however, it should be clear that in some
vacua a generalization of the discrete R-symmetry arguments of~\cite{Dine:1988kq} should be sufficient to
show existence of flat Higgs directions at least for special values of the complex structure moduli.  More generally, in vacua with a GLSM description one
can try to argue as follows.  To all orders in sigma model perturbation theory the possible F-term obstructions
are due to cubic couplings of the form $S\rep{27}\cdot\brep{27}$, where $S$ is some
bundle singlet~\cite{Dine:1988bz,Dine:1988kq}, and in vacua with a GLSM description these couplings are
strongly constrained~\cite{Berglund:1995yu}.  Thus, spacetime arguments may rule out or at least constrain
the possible F-term obstructions.   Of course once the $\alpha'$-perturbative obstructions have been shown to vanish,  one can reap the real benefit of the GLSM embedding by constraining or eliminating all together the worldsheet instanton effects that could lift the purported vacuum.

\section{A puzzle at the Landau-Ginzburg locus}
\label{s:LG}
In the previous section we argued that the M1 GLSM is a good description of the $\GE_6\to \SO(10)$ Higgs branch, at least in the neighborhood of the large radius limit.  Since this is the case,
given the spacetime arguments of~\cite{Dine:1988kq} and the GLSM worldsheet stability arguments of~\cite{Silverstein:1995re,Basu:2003bq,Beasley:2003fx}, it would be very surprising if M1
were not a sensible model at the Landau-Ginzburg (LG) locus --- the limit $r \to -\infty$.  Yet precisely this puzzling feature
was noted in~\cite{Distler:1993mk}:  the spectrum, as obtained by LG orbifold techniques, is not compatible with expectations based on the supersymmetric Higgs mechanism.  This surprising observation was a primary motivation for our
study, and in this section we will describe what we believe to be the resolution of the puzzle.
Before resolving the puzzle, our first goal will be to state it clearly. We will then gain some insight
by a mirror computation and describe the resolution.

\subsection{Massless spectrum at the Landau-Ginzburg locus}
\label{ss:LGquintic}
Massless spacetime fermionic states in a (0,2) heterotic compactification arise as right-moving
Ramond ground states and hence can be identified with elements of $\cH_{\cQb}$ --- the cohomology
of the $\cQb$ supercharge.  If the vacuum has a (0,2) GLSM description,
then barring accidents in the IR, we can hope to identify $\cH_{\cQb}$ of the
SCFT with the $\cQb$ cohomology of the GLSM.  Since the GLSM is a well-behaved
super-renormalizable theory,  one might hope that the cohomology computation is
reasonably tractable.

In order to identify massless states it is not sufficient to describe
$\cH_{\cQb}$;  in both the (NS,R) and (R,R) sectors one must also know the
left-moving quantum numbers, namely the energy $E$, the $\GUL$ charge $q$, and
representation of the linearly realized $\SO(10)$.\footnote{Since all of the matter states
are neutral under the hidden $\GE_8$, we will ignore its quantum numbers.}
Fortunately, it is possible to identify GLSM operators $T$ and $J$ in $\cH_{\cQb}$ that generate
a left-moving Virasoro$\times \GUL$ algebra~\cite{Silverstein:1994ih,Silverstein:1995re},
and these can be used to compute the requisite left-moving quantum numbers.  Moreover,
the GSO projection relates $\SO(10)$ representations to the $\GUL$ charges, while $\GUR$
charges, denoted by $\qb$, distinguish the types of spacetime supermultiplets~\cite{Kachru:1993pg}.

The computations are greatly simplified when the effects of the GLSM gauge instantons
are suppressed by going deep into the interior of a well-behaved phase.  In the M1 model
there are two limits where gauge instantons are suppressed:  (i) the large radius limit $r\to \infty$,
or (ii) the LG-locus $r \to -\infty$.  In the latter case, the excitations of $\Phi^0$, $\Sigma$ and $\Gamma^0$
are very massive, and the large VEV $|\phi_0|^2 = -5 r$ necessary to solve the
GLSM D-term Higgses $\GUG$ to $\Z_5$.  The remaining light degrees
of freedom, $\Phi^i$, $\Gamma^i$ and $\Gamma^6$ are described by a Landau-Ginzburg
orbifold with chiral (0,2) superpotential
\begin{align}
\cW_{\text{LG}} = \Gamma^i J_i(\Phi) + \Gamma^6 J_6(\Phi),
\end{align}
and $\GUL\times\GUR$ charges listed in table~\ref{t:LGM1}.
\begin{table}
\begin{center}
\begin{tabular}{|c|c|c|c|c|}
\hline
$~$ 		&$\theta$		&$\Phi^i$	 	&$\Gamma^i$	&$\Gamma^6$  \\ \hline
$\GUL$ 	&$0$			&$\ff{1}{5}$	&$-\ff{4}{5}$	&$-1$  \\ \hline
$\GUR$ 	&$1$			&$\ff{1}{5}$	&$\ff{1}{5}$	&$0$  \\ \hline
\end{tabular}
\caption{M1 symmetries at the LG locus.}
\label{t:LGM1}
\end{center}
\end{table}
The generator of the $\Z_5$ gauge symmetry acts on the fields by $e^{2\pi i q}$.

Many properties of such LG orbifolds are reasonably well-understood both in
the context of type II compactifications~\cite{Vafa:1989xc,Intriligator:1990ua},
and heterotic vacua~\cite{Kachru:1993pg,Distler:1993mk}.  For our purposes,
the most important simplification obtained at the LG locus is the computation of
$\cH_{\cQb}$,  which may be accomplished in two steps:
restrict to right-moving zero modes and represent $\cQb_{\text{LG}}$ on
the remaining excitations via
\begin{align}
\cQb_{\text{LG}} = \oint \frac{dz}{2\pi i} \left( \gamma^i J_i (\phi) + \gamma^6 J_6 (\phi) \right).
\end{align}
Note that we have implicitly rotated to Euclidean signature, and we will find it convenient to
take our worldsheet to be the plane.

The action of the $\Z_5$ gauge symmetry can be conveniently combined with the GSO projection
by introducing twisted ground states $|k\ra$, $k=0,\ldots, 9$, where the internal fields have periodicities shifted by $e^{i\pi  q k}$, while the $8$ free fermions $\xi$ are anti-periodic for $k$ even and periodic
for $k$ odd.  The GSO projection is then carried out as follows. In NS sectors ($k$ odd), we project
onto states with $e^{-i\pi J} (-)^{F_\xi} = 1$; in R sectors ($k$ even) states with $q$ odd pair up with the $\rep{8}^s$ twist fields of the $\xi$ system, while those with $q$ even are paired
with the $\rep{8}^c$ twist fields.  Finally, since we are interested in massless states, level
matching allows us to restrict attention to states with total left-moving energy zero.

With these ingredients in hand, we have a simple algorithm to compute the massless
spectrum~\cite{Kachru:1993pg,Distler:1993mk}:
\begin{itemize}
\item[1.] compute the quantum numbers $E$, $q$, and $\qb$ of the twisted vacua $|k\ra$   (general expressions for
these in LG orbifolds can be found in~\cite{Distler:1993mk});
\item[2.] construct the $E=0$ states by acting on $|k\ra$ with lowest raising modes of the fields in each
sector, and project onto appropriate values of $q$;
\item[3.] compute $\cH_{\cQb}$;
\item[4.] identify spacetime multiplets as follows:  a state with $\qb = -1/2$ belongs to a chiral matter multiplet; one with $\qb = 1/2$ is in an anti-chiral multiplet; states with $\qb = \pm 3/2$ are gauginos
in vector multiplets, with $\qb = 3/2$ being right-handed.
\end{itemize}
The computation is further simplified by noting that CPT exchanges sector $k$ with $10-k$ for $k >0$, so that we need only consider $k=0,\ldots, 5$.

We can recover the $\GE_6$ locus of the quintic theory by setting $J_6 = 0$ and keeping $J_i$ generic.  In this case, the free $\Gamma^6$ can be treated as part of the $\xi$ system, and
instead of labeling states by their $\SO(8)\times \GUL$ quantum
numbers, it is more natural to use $\SO(10)\times\GUL'$
labels.\footnote{Note that the twisted vacua $|2k\ra$ have $q' = q+1/2$,
while $|2k+1\ra$ vacua have $q' = q$.}  The latter labeling
yields the same description as originally obtained in~\cite{Kachru:1993pg}.   As we
are interested in deformations that break $\GUL'$ and preserve $\GUL$, it will
be more useful to work with the $\SO(8)\times\GUL$ representations described
above; however for convenience and comparison, we will list both representations.

Applying the algorithm and concentrating on states with $\qb < 0$, we find the following massless fermions.\footnote{The gravitino, dilatino, and the hidden $\GE_8$ gauginos may all be found in the $k=1$ sector.}
\begin{itemize}
\item[1.] Gauginos $\rep{78} = \rep{16}_{-3/2} \oplus \rep{45}_0\oplus \rep{1}_0 \oplus \brep{16}_{3/2}.$
\begin{align*}
 \rep{16}_{-3/2} & = \rep{8}^c_{-2} \oplus \rep{8}^s_{-1} & (k=0), \nonumber\\
\rep{45}_0\oplus \rep{1}_0& = \rep{28}_{0}\oplus \rep{8}^v_{-1}
\oplus \rep{8}^v_{1} \oplus \rep{1}_{0}^{\oplus 2}  & (k=1), \nonumber\\
\brep{16}_{3/2} & = \rep{8}^c_{2}\oplus \rep{8}^s_{1} & (k=2).
\end{align*}
\item[2.] Matter $\rep{27} = \rep{10_{-1}} \oplus \rep{16}_{1/2}\oplus \rep{1}_2$.
\begin{align*}
\rep{10}_{-1} & = \rep{8}^v_{-1} \oplus \rep{1}_{-2} \oplus \rep{1}_{0}  & (k=3), \nonumber\\
\rep{16}_{1/2} & = \rep{8}^c_{0} \oplus \rep{8}^s_{1} & (k=4), \nonumber\\
\rep{1}_{2} & = \rep{1}_{2} & (k=5).
\end{align*}
\item[3.] Matter $\brep{27} = \rep{1}_{-2} \oplus \brep{16}_{-1/2} \oplus \rep{10}_1$.  ($101$ of these.)
\begin{align*}
\rep{1}_{-2} & = \rep{1}_{-2} & (k=9), \nonumber\\
\brep{16}_{-1/2} & = \rep{8}^s_{-1} \oplus \rep{8}^c_{0}  & (k=0), \nonumber\\
\rep{10}_1 & = \rep{8}^v_{1} \oplus \rep{1}_{0} \oplus \rep{1}_{2} & (k=1).
\end{align*}
\item[4.] Neutral matter.  Finally, we have $\rep{1}^{\oplus 301}_{0}$ from $k=1$ and $\rep{1}^{\oplus 25}_{0}$ from $k=3$.
\end{itemize}

Special values of $J_i$ can lead to additional massless states associated with an enhanced abelian symmetry.  For instance, by tuning to the Gepner values $J_i = \Phi_i^4$ we find that the $k=1$
sector contains four more gauginos and four more $\GE_6$-neutral singlets.  As we move away
from such special points the extra gauginos and matter are paired up by the Higgs mechanism.  This is manifested in the LG computation by a change in $\cH_{\cQb}$.  In particular, the
zero energy $q=0$ states in the $k=1$ sector take the form
\begin{align}
\label{eq:basicH}
\xymatrix{ 0 \ar[r] & \C^{25} \ar[r]^{\cQb} &\C^{350} \ar[r] & 0 \\
               \qb =-\ff{5}{2} & \qb=-\ff{3}{2} & \qb = -\ff{1}{2}
               & \qb = \ff{1}{2} }.
\end{align}
When $J_i$ is generic, $\cQb$ has a one-dimensional kernel at $\qb=-3/2$; for non-generic choices of $J_i$ the dimension of the kernel jumps. For example, at the Fermat point the kernel becomes $5$-dimensional, leading to four more gauginos and four more chiral fermions.

\subsection{The puzzle} \label{ss:LGpuzzle}
Having reviewed the massless spectrum computation at the LG locus with $J_6=0$, we now describe the modifications when
$J_6 \not\in \la J_1,\ldots, J_5\ra$.  The naive expectation is that turning
on $J_6$ corresponds to  giving VEVs to the spacetime scalar fields $\phi = \rep{1}_{0} \in \rep{27}$ and $\phit = \rep{1}_0 \in \brep{27}$.  Thus, we expect that some of the fermions will get masses by D-terms and others via F-terms.\footnote{In this section the $\lambda$ is an $\GE_6$ gaugino, while $\psi$($\psit$) is a fermion in the chiral $\rep{27}$($\brep{27}$) matter multiplet.}
First, the gauginos and charged matter fermions should be paired up
by the gauge Yukawa terms, with mass terms of the form
\begin{align}
\cL^{D}_{\text{Yuk} } &= i \phi^\dag \left\{
\lambda_{\rep{8}^s_{-1}} \cdot \psi_{\rep{8}^s_1} +
\lambda_{\rep{8}^v_1} \cdot \psi_{\rep{8}^v_{-1}} \right\}
+ i\phit^\dag \left\{
\lambda_{\rep{8}^s_1} \cdot \psit_{\rep{8}^s_{-1}} +
\lambda_{\rep{8}^v_{-1}} \cdot \psit_{\rep{8}^v_1} \right\} \nonumber\\
~&~~+i \lambda_{\rep{1}_0}  \left\{\phi^\dag \psi_{\rep{1}_0} -\phit^\dag \psit_{\rep{1}_0} \right\}
+\text{h.c.} ~.
\end{align}
In the LG description of the spectrum this can only be manifested
by a change in the cohomology akin to that described below~(\ref{eq:basicH}).  That is, turning on $J_6$ should lead to a $\Delta \cQb$, which must provide new non-zero maps among the states. Specifically, to implement the spacetime Higgs mechanism, $\Delta\cQb$ should lead to the following non-zero
maps (we now specify the $q$ and $\qb$ charges of the states)
\begin{align}
\label{eq:presentmapD}
\Delta\cQb~~:~~
 \underbrace{\rep{8}^s_{-1,-3/2} \longrightarrow \rep{8}^s_{-1,-1/2}}_{k=0} ,\quad
\underbrace{ \rep{8}^v_{1,-3/2} \longrightarrow \rep{8}^v_{1,-1/2}, \qquad
 \rep{1}^{\oplus 2}_{0,-3/2} \longrightarrow \rep{1}_{0,-1/2}}_{k=1},
\end{align}
as well as
\begin{align}
\label{eq:absentmapD}
\Delta\cQb ~~:~~&
\underbrace{\rep{8}^s_{1,-3/2}}_{k=2} \longrightarrow
\underbrace{\rep{8}^s_{1,-1/2}}_{k=4}, \qquad
\underbrace{\rep{8}^v_{-1,-3/2}}_{k=1} \longrightarrow
\underbrace{\rep{8}^v_{-1,-1/2}}_{k=3}, \cr
&
\underbrace{ \rep{1}^{\oplus 2}_{0,-3/2}}_{k=1} \longrightarrow
\underbrace{ \rep{1}_{0,-1/2}}_{k=3}.
\end{align}

The second change in the massless spectrum is due to the F-terms.   For instance,
the $\rep{27}^3$ and $\brep{27}^3$ couplings lead to the following F-term mass terms
(for simplicity we suppress the indices on the $\brep{27}$):
\begin{align}
\cL^F_{\text{Yuk}} =
\phi  (\psi_{\rep{8}^{c}_0} \cdot \psi_{\rep{8}^{c}_0} + \psi_{\rep{1}_{-2}} \psi_{\rep{1}_{2}})
+\phit  (\psit_{\rep{8}^c_0} \cdot \psit_{\rep{8}^c_0} + \psit_{\rep{1}_{-2}} \psit_{\rep{1}_{2}}) +
\text{h.c.}~.
\end{align}
In the LG description, these should lead to masses for the $102$ $\rep{10}$s via the following maps:
\begin{align}
\label{eq:presentmapF}
\Delta\cQb ~~:~~ \underbrace{\rep{8}^{c \oplus 101}_{0,-1/2} \longrightarrow \rep{8}^{c\oplus 101}_{0,1/2}}_{k=0}, \qquad
 \underbrace{\rep{1}^{\oplus 101}_{2,-1/2} \longrightarrow \rep{1}^{\oplus 101}_{2,1/2}}_{k=1},
\end{align}
and
\begin{align}
\label{eq:absentmapF}
\Delta\cQb ~~:~~ \underbrace{\rep{8}^c_{0,-1/2}}_{k=4} \longrightarrow
             \underbrace{\rep{8}^c_{0,1/2}}_{k=6}, \qquad
\underbrace{\rep{1}_{-2,-1/2}}_{k=3} \longrightarrow
\underbrace{ \rep{1}_{-2,1/2}}_{k=5}.
\end{align}
Note that the $\rep{1}_{2,1/2}$ state in $k=1$ is CPT conjugate to the $\rep{1}_{-2,-1/2}$, $k=9$ state quoted in the $\brep{27}$ decomposition, and similarly the $\rep{1}_{-2,1/2}$ in $k=5$ is
CPT conjugate to the $\rep{1}_{2,-1/2}$ quoted in the $\rep{27}$
decomposition.  Provided that all of these maps are non-trivial, we would find a sensible
spectrum of an $\SO(10)$ theory, with massless spectrum consisting of the following $\qb<0$
states:
\begin{itemize}
\item[1.] gauginos: $
\rep{45}  = \underbrace{\rep{8}^c_{-2}}_{k=0}\oplus
\underbrace{ \rep{28}_{0}\oplus  \rep{1}_{0}}_{k=1} \oplus
\underbrace{\rep{8}^c_2}_{ k=2}$;
\item[2.] $\SO(10)$-charged matter:
$\rep{16}^{\oplus 100} = \underbrace{\rep{8}^{s\oplus 100}_{-1}}_{k=0}\oplus\underbrace{\rep{8}^{v\oplus 100}_1}_{k=1}$;
\item[3.] neutral matter, consisting of $427$ states with components in $k=1$ and $k=3$
sectors.
\end{itemize}
Carrying out the computation for generic $J_i$ and $J_6$, we find
all of these states and explicitly identify the maps in~(\ref{eq:presentmapD}) and~(\ref{eq:presentmapF}).\footnote{Details of the computation are provided in appendix~\ref{app:LG}.}
For special values of the $J$ the $\brep{27}^3$ couplings can develop zeroes, leading to the vanishing of some of the maps in~(\ref{eq:presentmapF}) and therefore to additional massless $\rep{10}$ fields.  For instance,  setting $J_i = \phi_i^4$ and $J_6 = \phi_1\phi_2\phi_3\phi_4\phi_5$, we find $50$ massless $\rep{10}$s, in  agreement with the large radius computation~\cite{Donagi:2006yf}.  These states are also accompanied by an enhanced $\GU(1)^4$ gauge symmetry with corresponding massless scalars.

Unfortunately, the sensible $\SO(10)$ spectrum is accompanied by extra massless states.  Their origin is simple
to understand: since  $\Delta\cQb$
manifestly preserves the twisted vacuum, it cannot lead to
the maps in~(\ref{eq:absentmapD}) or in~(\ref{eq:absentmapF}).
Thus, the fermions in~(\ref{eq:absentmapD}) and~(\ref{eq:absentmapF}) remain massless.  From the spacetime point of view, it appears that we are working in a ``vacuum'' with $\phi = 0$ and $\phit \neq 0$;
but that is not consistent with $N=1$ spacetime supersymmetry!

It should be clear that this conundrum is not confined to the quintic.  In fact, the Higgs
deformations of any (2,2) vacuum with a GLSM description and a Landau-Ginzburg
locus will have the same sort of paradoxical spectrum.
This is the puzzle we wish to resolve.

At this point it is good to remember that starting at the
Gepner point we can give a VEV to $\phi$ by deforming
the SCFT by the bosonic twist field that is the superpartner of the $\rep{1}_0$ fermion in the $k=3$ sector.  By the arguments
of~\cite{Dine:1988kq}, we are guaranteed a marginal
direction with $\GE_6$ broken to $\SO(10)$, provided we
tune the $\phi$ and $\phit$ VEVs appropriately.
Of course the deformation by a twist field necessarily breaks
the quantum symmetry of the LG orbifold~\cite{Vafa:1989ih,Distler:1994hs} and thus cannot be described as a change
in the $\cQb$-cohomology of the LG theory.

We also recall that the original M1 GLSM did encode both the $\rep{27}$ and
$\brep{27}$ infinitesimal deformations, and before reduction to the LG locus
that theory certainly has
effects that break the quantum symmetry of the orbifold --- namely the gauge instantons.  Thus,
it is reasonable to guess that the problem lies in the reduction of the GLSM to the LG description.  To explore this guess it would be useful to study the effect of the $\ep$ coupling for finite $r$; although this is challenging to do directly, it is relatively straightforward in the mirror description, to which we turn next.

\subsection{A glance in the mirror}
The mirror LG orbifold for the quintic is obtained by supplementing the $\Z_5$ quotient by an
additional $\Z_5^3$ symmetry, which acts on the $\Phi^i,\Gamma^i$ by
\begin{align}
(\Phi^i, \Gamma^i) \to e^{2\pi i t^a w^a_i} (\Phi^i, \Gamma^i),
\end{align}
where $t^a =0,\ldots,4$ for $a=1,2,3$, and the generators $w^a_i$ can be taken
as\footnote{We are indebted for this choice of basis to~\cite{Candelas:1990qd}.}
\begin{align}
w^1  = (0, \ff{1}{5}, 0, 0, -\ff{1}{5}), \quad
w^2  = (0, 0, \ff{1}{5},  0, -\ff{1}{5}), \quad
w^3  = (0, 0, 0, \ff{1}{5},  -\ff{1}{5}).
\end{align}
On the $\GE_6$ locus, the most general (0,2) potential compatible with the
$\Z_5\times\Z_5^3$ orbifold symmetry is
\begin{align}
\label{eq:LGpot}
\cW = \Gamma^i J_i, \qquad J_i = \Phi_i^4 - \psi \prod_{j\neq i} \Phi_j.
\end{align}
Turning on the $\rep{27}$ VEV in the original theory is equivalent to turning on a $\brep{27}$ VEV in the
mirror, and following the construction of section~\ref{ss:quindef}, we see that the candidate GLSM operator
is uniquely determined as
\begin{align}
\Delta \cW = z \Gamma^6 \prod_i \Phi_i.
\end{align}
The computation of the massless fermion spectrum proceeds essentially as before.  The new complications are the additional twisted sectors of the $\Z_5^3$ orbifold and the requisite projection onto $\Z_5^3$-invariant states.  It is important that the twisted vacua $|k;\bt\ra$ carry $\Z_5^3$ charges, which may be computed as in~\cite{Aspinwall:2010ve}.

First setting $z=0$, we reproduce the usual quintic spectrum, but this time in the mirror
description.  When $\psi \neq 0$, we find the following states with $\qb <  0$.  As before,
we give both the $\SO(10)\times\GUL'$ and $\SO(8)\times \GUL$ decompositions.
In the former case the free left-moving fermions are $\xi^\alpha$, $\alpha =1,\ldots, 10$, while
in the latter they are $\xi^a$, $a=1,\ldots, 8$.  By a slight abuse of notation we refer by the
same name to the field and its lowest excited (possibly zero) mode in each sector.  We also
define $\mon = \prod_i \phi_i$ and its conjugate $\monb = \prod_i \phib_i$. More details on our LG conventions can be found in appendix~\ref{app:LG}.
\begin{itemize}
\item[1.] Gauginos, $\rep{78} = \rep{16}_{-3/2} \oplus \rep{45}_0\oplus \rep{1}_0 \oplus \brep{16}_{3/2}.$
\begin{align*}
|0;\bzer\ra \leftrightarrow \rep{16}_{-3/2}  &
\begin{cases}
~~~|0;\bzer\ra \leftrightarrow  &\rep{8}^c_{-2}\\
\gammab^6 |0;\bzer\ra \leftrightarrow &\rep{8}^s_{-1}.
\end{cases}  \nonumber \\
\xi^\alpha \xi^\beta |1;\bzer\ra \leftrightarrow  \rep{45}_0 &
\begin{cases}
\xi^a \xi^b |1;\bzer\ra \leftrightarrow &\rep{28}_0 \\
\xi^a \gamma_6 |1;\bzer\ra\leftrightarrow &\rep{8}^v_{-1}  \\
\xi^a \gammab_6 |1;\bzer\ra\leftrightarrow  &\rep{8}^v_{1} \\
\gammab_6\gamma_6 |1;\bzer\ra  \leftrightarrow &\rep{1}_0
\end{cases} \nonumber\\
{\textstyle \sum_i} (\phi_i\phib_i+4\gamma_i\gammab_i) |1;\bzer\ra \leftrightarrow  \rep{1}_0& \nonumber\\
|2;\bzer\ra \leftrightarrow \brep{16}_{3/2}  &
\begin{cases}
~~~|2;\bzer\ra \leftrightarrow  &\rep{8}^s_{1}\\
\gammab^6 |2;\bzer\ra \leftrightarrow &\rep{8}^c_{2}
\end{cases}
\end{align*}
\item[2.] Matter $\brep{27} = \rep{1}_{-2} \oplus \brep{16}_{-1/2} \oplus \rep{10}_1$.
\begin{align*}
\monb^2 |9;\bzer\ra \leftrightarrow \rep{1}_{-2} &\nonumber\\
\mon |0;\bzer\ra \leftrightarrow \brep{16}_{-1/2} &
\begin{cases}
\gammab^6 \mon  |0;\bzer\ra  \leftrightarrow  & \rep{8}^c_{0} \\
~~~\mon |0;\bzer\ra \leftrightarrow & \rep{8}^s_{-1}
\end{cases} \nonumber\\
\xi^\alpha \mon |1;\bzer\ra \leftrightarrow \rep{10}_1 &
\begin{cases}
\gammab_6 \mon|1;\bzer\ra\leftrightarrow & \rep{1}_{2}   \\
\xi^a \mon |1;\bzer\ra\leftrightarrow  & \rep{8}^v_1  \\
\gamma_6 \mon |1;\bzer\ra \leftrightarrow& \rep{1}_0
\end{cases}
\end{align*}
\item[3.] Matter $\rep{27} = \rep{10}_{-1} \oplus \rep{16}_{1/2}\oplus \rep{1}_2$.
There are $101$ of these, liberally scattered through the various $|k;\bt\ra$ twisted sectors.  Fortunately, we will not need their explicit form. 
\item[4.] Neutral matter.  There are $326$ of these, also scattered throughout the twisted sectors.
\end{itemize}
When $\psi = 0$, the only modification to the spectrum is the appearance of four more
gauginos in the $|1;\bzer\ra$ sector.   The $\sum_i (\phi_i\phib_i+4\gamma_i\gammab_i) |1;\bzer\ra$ gaugino is replaced by
$\oplus_i (\phi_i\phib_i+4\gamma_i\gammab_i) |1;\bzer\ra$, which are accompanied by four additional $\GE_6$-neutral singlets at $\qb = -1/2$.

Having described the spectrum for $z=0$, we can now turn on a $J_6$ deformation
and study the modifications due to
\begin{align}
\Delta \cQb =  \gamma_6 J_6 + \gammab_6^\dag J_6'.
\end{align}
The $\gammab_6^\dag$ is the conjugate mode to the first excited mode of $\gammab_6$,
and $J_6$ and $J_6'$ are quintic polynomials obtained by expanding the operator $z \mon$
in the modes $\phi_i$ and $\phib_i^\dag$.  Since $\Delta\cQb$ does not change the left-moving
energy, some or all of these terms are zero in most of the twisted sectors.  In fact, $\Delta \cQb$ only leads to modifications in the untwisted sector, the $|1;\bzer\ra$ sector, and its CPT conjugate $|9;\bzer\ra$.

In the untwisted sector we reduce to zero modes, and $\Delta \cQb = z\gammab_6^\dag \mon$ leads to the following non-trivial maps:
\begin{align}
\Delta\cQb ~~:~~&\gammab_6 |0;\bzer\ra \longrightarrow \mon |0;\bzer\ra, \quad
\mon \gammab_6 |0;\bzer\ra \longrightarrow \mon^2|0;\bzer\ra, \cr
&\mon^2 \gammab_6 |0;\bzer\ra \longrightarrow \mon^3|0;\bzer\ra.
\end{align}
The last is CPT conjugate to the first, and matching the charges
to the $\xi^a$ spin fields, we find the maps
\begin{align}
\label{eq:mDQ1}
\Delta\cQb~~:~~
&\rep{8}^s_{-1,-3/2} \longrightarrow \rep{8}^s_{-1,-1/2}, \qquad
\rep{8}^c_{0,-1/2} \longrightarrow \rep{8}^c_{0,1/2}, \cr
&\rep{8}^s_{1,1/2} \longrightarrow \rep{8}^s_{1,3/2}.
\end{align}
All of the untwisted states are lifted, with the exception of
the gauginos $\rep{8}^c_{-2,-3/2} \leftrightarrow |0;\bzer\ra$
and their CPT conjugates $\rep{8}^c_{2,3/2} \leftrightarrow \gammab^6 \mon^3 |0;\bzer\ra$.

In the $|1;\bzer\ra$ sector we find (for brevity we suppress the ket $|1;\bzer\ra$):
\begin{align}
\Delta\cQb~~:~~\xi^a \gammab_6  \longrightarrow \xi^a\mon, \qquad
\gamma_6\gammab_6\oplus\phi_i\phib_i\longrightarrow
\gamma_6\mon, \qquad
\gammab_6 \mon \longrightarrow \mon^2,
\end{align}
which corresponds to
\begin{align}
\label{eq:mDQ2}
\Delta\cQb ~~:~~&\rep{8}^v_{1,-3/2} \longrightarrow \rep{8}^v_{1,-1/2},\qquad
\rep{1}^{\oplus 2}_{0,-3/2} \longrightarrow \rep{1}_{0,-3/2}, \cr
&\rep{1}_{2,-1/2} \longrightarrow \rep{1}_{2,1/2}.
\end{align}
Comparing~(\ref{eq:mDQ1}) and~(\ref{eq:mDQ2}) to~(\ref{eq:absentmapD}) and~(\ref{eq:absentmapF}), and remembering
to change $q\to-q$, we see that mirror symmetry predicts exactly the puzzling missing maps of the original LG computation.

\subsection{Puzzle resolution via the mirror map}
Having assured ourselves that all of the expected mass terms
are indeed generated for $z\neq 0$, we are now ready to explain
the puzzle in the original description.

In order to connect the original M1 description with its mirror,
we need to map the operators $\cO_a$ and $\Oh_m$ of~(\ref{eq:M1def}) to the operators in the mirror theory.  This is easily
done for the ``toric'' $\cO_a$ and ``polynomial''
$\Oh_m$ via the monomial-divisor mirror
map~\cite{Aspinwall:1993rj,Morrison:1994fr}.  In the
case of the quintic the result is simple:  the M1 GLSM
operator $\sigma$ is identified with the monomial $\mon$ in the mirror theory.

The crucial point is that the operators $\sigma$ and $\mon$
correspond to infinitesimal deformations of the complexified K\"ahler
moduli space of the quintic, parametrized by $\tau$ in the original
model and by $\psi$, defined in~(\ref{eq:LGpot}), in the mirror
description.   Since the chiral ring elements are to be identified with the
cotangent space to the moduli space, the operators $\sigma$ and $\mon$
should be identified as
\begin{align}
\sigma  =    \frac{d \psi}{d \tau} \mon.
\end{align}
This is perhaps familiar from the relation of the three-point functions
\begin{align}
\la \sigma^3\ra_{\text{quintic}} = \left(\frac{d \psi}{d \tau}\right)^3 \la \mon^3 \ra_{\text{mirror}}.
\end{align}
The monomial-divisor mirror map identifies $e^{2\pi i\tau} = (-5\psi)^{-5}$,
 so the M1 GLSM deformation corresponding to turning on the
$\rep{27}$ VEV is mapped as follows:
\begin{align}
\xymatrix{
\ep \gammab^6 \cQ_{-1/2}\cdot \sigma  \ar[rr]^{\text{mirror}}&&
z \gamma^6 \cQ_{-1/2} \cdot \mon },
\end{align}
with $$z(\ep, q) = \ep  \frac{ d\psi}{d \tau} \sim \ep e^{-2\pi i \tau/5}.$$
In order to reach the LG locus of the M1 model, we know that
we must tune $r \to -\infty$, but what shall we do with the other
parameters?  The simplest possibility is to keep them fixed.
In this case the $\cQb$ operator of the GLSM reduces to that
of the LG theory, and we obtain the ``puzzling'' spectrum.
In the mirror description this sets $z = 0$, thereby eliminating
the non-trivial $\Delta\cQb$ maps identified above.
Alternatively, as we take $r \to -\infty$ we can scale
$\ep \sim e^{2\pi i\tau/5}$, in which case $z$ stays finite, and
the mirror computation produces the expected maps.
However, there is a price to pay in the original model:
since some of the GLSM parameters are now getting parametrically
large as $r \to -\infty$,  $\cQb_{\text{GLSM}}$ need not reduce to
$\cQb_{\text{LG}}$.  In fact, it is quite natural to expect that
$\cQb_{\text{GLSM}}$ has a gauge instanton expansion, which
in the LG phase takes the form
\begin{align}
\cQb_{\text{GLSM}} = \cQb_{\text{LG}} + e^{-2\pi i\tau/5} \cQb_{1} + \cdots.
\end{align}
A comparison with gauge instantons in the large radius
phase, which yield corrections proportional to $e^{2\pi i \tau n}$, $n \in \Z_{\ge 0}$,  suggests that the factor $e^{-2\pi i \tau/5}$ be interpreted as a fractional instanton effect.  As explained in~\cite{Witten:1993yc} this is the
GLSM avatar of an insertion of a $\Z_5$ twist field in the orbifold theory.   In fact, $\cQb_1$ will map states in the vacuum $|k\ra$ to those in $|k+2\ra$, which is exactly what the
``missing maps'' are supposed to do.

This is our proposed resolution: in order to turn on a $\rep{27}$ VEV in the LG limit of the M1 GLSM, the natural normalization
of the GLSM operators requires us to scale $\ep \sim e^{2\pi i \tau/5}$ as we take
the $r\to -\infty$ limit.  This leads to an unsuppressed gauge instanton correction
to $\cQb_{\text{LG}}$ which modifies the spectrum.  It would be very interesting to compute $\cQb_1$ directly in the GLSM, but we will not pursue it in this paper.

Finally, it should be noted that this phenomenon has a clear echo
in the spacetime theory.  The charged matter kinetic terms have
a non-trivial dependence on both types of moduli~\cite{Dixon:1989fj}, with metrics for the $\rep{27}$ and $\brep{27}$ fields given by,
respectively,
\begin{align}
G = g~ e^{ (\Kh -K)/3}, \qquad
\Gh= \gh~ e^{(K-\Kh)/3}.
\end{align}
The $K$ and $\Kh$ are K\"ahler potentials for the complexified K\"ahler and complex structure moduli spaces, while the $g$ and $\gh$ are the corresponding metrics.  In order to obtain sensible results for the Higgs mechanism one must work with properly normalized matter fields.  Although the LG locus is at finite distance in the moduli space, in the natural GLSM normalization of the operators, as one takes the $r\to \infty$ limit, the $\rep{27}$ kinetic terms are driven to zero, while the $\brep{27}$ ones are driven to infinity.

\section{Outlook}
In this note we made a small step in exploring the Higgs branch in $N=1$ $d=4$ compactifications of the heterotic string.  We argued that a GLSM description of
a heterotic vacuum with (2,2) worldsheet supersymmetry naturally includes deformations
that correspond, at least at the infinitesimal level, to $\GE_6$-breaking deformations.
It is clear that in a large set of examples this construction will yield exactly marginal
deformations, but the precise conditions under which this should be true need to be clarified.
One could, for instance, try to directly prove stability of bundles defined by
generalizing~(\ref{eq:M1bundle}) to a Calabi-Yau hypersurface in a toric variety.

The identification of the natural GLSM deformations corresponding to $\rep{27}$ and $\brep{27}$
VEVs relied on a two simple facts, which are probably good to keep in mind if one is interested in
the construction's generalizations. First, we used the relation between the $\rep{10}_{\pm 1}$ charged
matter vertex operators and (2,2) moduli.  Second,
we restricted attention to the (2,2) moduli that can be realized as deformations of the GLSM Lagrangian.
With these provisos, however, the form of the GLSM deformations is reasonably clear.

By using mirror symmetry, we argued that the paradoxical spectrum of the deformed theory at the
Landau-Ginzburg locus is an artifact of the operator normalizations natural in the
``algebraic gauge'' coordinates~\cite{Morrison:1994fr}  of the GLSM.  While this does account for
the puzzle, it would be more satisfactory and likely instructive to re-examine the reduction
of the original GLSM and directly compute the gauge instanton correction to $\cQb_{\text{LG}}$.
We believe this should be doable.  This computation may also cast light on the relationship between
M1 and M2 GLSMs.

Given a vacuum with flat Higgs deformations encoded in the GLSM, there are many questions
one can explore.  The GLSM parameter space modulo field redefinitions
should yield an algebraic description of this (0,2) moduli space as a holomorphic quotient.
What sorts of singularities are encountered?   To what extent can we continue to picture the
moduli space as splitting into K\"ahler, complex structure, bundle, and Higgs moduli?  Can we
generalize the quintic example and construct a mirror map relating the mirror Higgs deformations?

Perhaps the most striking difference between the Higgs deformations and deformations by
the gauge-neutral moduli is the discontinuous behavior of the topological heterotic ring identified
in~\cite{Adams:2003zy,Katz:2004nn,Adams:2005tc}.  Given the broad conditions under which
this structure has been shown to exist, it might be at first surprising that it should behave
discontinuously under small deformations of the theory.  However, the reason is readily found:
the Higgs deformations, as opposed to the gauge-neutral deformations, break the $\GUL'$
symmetry of the (0,2) SCFT.  It will be interesting to explore the possible discontinuities and
the distinct topological heterotic rings that are realized on different loci in the same heterotic
moduli space.  The GLSM description is likely to be our best tool for these explorations.

{\bf Acknowledgments}: This project was initiated in a discussion with
J.~Distler, J.~Lapan, and R.~Plesser at the BIRS workshop on (0,2)
mirror symmetry in March 2010, and we are happy to thank BIRS
for providing the location and logistics, and the participants for
providing a fun and stimulating atmosphere.  IVM would like to thank
R.~Plesser for quite a number of further useful discussions of this project,
as well as spotting an error in an early version of the manuscript.
IVM also thanks S.~Hellerman for discussions on discrete
symmetries that led to an improved appendix~\ref{app:permsym}.
We have also benefited from the hospitality of our respective institutions
while this work was being completed.  JM is supported by an
EPSRC Postdoctoral Fellowship EP/G051054/1, and IVM is supported in
part by the German-Israeli Project cooperation (DIP H.52) and the
German-Israeli Fund (GIF).

\begin{appendix}

\section{Some details of the LG spectrum computation}
\label{app:LG}
\subsection{Orbifold generalities}
We follow the method of~\cite{Kachru:1993pg,Distler:1993mk}, mostly in the notation of~\cite{Aspinwall:2010ve}.  Given a (2,2) $(c,\cb)  = (9,9)$ theory defined by a superpotential $\cW = \Gamma^i P_{,i} (\Phi)$, with the $\GUL$ charges of the $\phi_i$ denoted
by $\alpha_i$, we are interested in the $\Z_{2d}$ orbifold  generated by
$e^{-i \pi J}$, under which
\begin{align}
\phi^i \mapsto e^{- i \pi k\alpha_i/2} \phi^i, \qquad
\gamma^i \mapsto e^{-i \pi k(\alpha_i-1)/2} \gamma^i.
\end{align}
The $k=0,\ldots 2d -1$ twisted sectors are defined by the following modings for the
holomorphic fields:
\begin{align}
\phi^i(z) &= \sum_{s\in \Z-\nu_i} \phi^i_s z^{-s-h_i},
&\qquad
& \gamma^i(z)  = \sum_{s\in \Z-\nut_i} \gamma^i_s z^{-s-\htld_i} \nonumber\\
\phib^i(z) &= \sum_{s\in\Z+\nu_i} \phib^i_s z^{-s+h_i-1},
&\qquad
&\gammab^i(z) = \sum_{s\in\Z+\nut_i} \gammab^i_s z^{-s+\htld_i-1},
\end{align}
where $2h_i = \alpha_i$, $2\htld_i = 1+\alpha_i$,
\begin{alignat}{2}
\label{eq:nus}
\nu_i  &= \frac{k\alpha_i}{2}\pmod 1 &\qquad 0&\le\nu_i<1,\nonumber\\
\nut_i &= \frac{k(\alpha_i-1)}{2} \pmod 1 &\qquad-1&<\nut_i\le 0.
\end{alignat}
We denote the lowest excited modes by
\begin{align}
\phi_i \equiv \phi^i_{-\nu_i}, \qquad
\phib_i \equiv \phib^i_{\nu_i-1}, \qquad
\gamma_i \equiv \gamma^i_{-1-\nut_i}, \qquad
\gammab_i \equiv \gammab^i_{\nut_i}.
\end{align}
With our conventions a twisted vacuum $|k\ra$ is annihilated by
$\phib_i$ ($\gamma_i$)  whenever $\nu_i = 0$ ($\nut_i = 0$).
We use a dagger to denote the conjugate lowering modes --- for instance
$\phi_i^\dag \equiv \phib^i_{\nu_i}$.
Since we are interested in the $\SO(10)$-preserving deformations, we
will also need to describe the modes of $\gamma^6$.  This is simple:
for even $k$ $\gamma^6$ has a zero mode and $\gamma_6 |k\ra =0$,
while for odd $k$ both $\gamma_6 = \gamma^6_{-1/2}$ and $\gammab_6 = \gammab^6_{-1/2}$ increase the left-moving energy $E$ by $1/2$.  In the conventions as above, the twisting of $\gamma^6$ is described by $\nut_6 =0$ for $k$ even and $\nut_6 = -1/2$ for
$k$ odd.

The quantum numbers of the twisted states are given by
\begin{align}
\label{eq:charges}
q &= \sum_i \left[ (\alpha_i-1) (\nut_i + \ff{1}{2}) -\alpha_i (\nu_i -\ff{1}{2}) \right] -(\nut_6+\ff{1}{2}), \nonumber\\
\qb & = \sum_i \left[ \alpha_i(\nut_i + \ff{1}{2}) +(1-\alpha_i) (\nu_i -\ff{1}{2}) \right],
\\
E & = \begin{cases}  0 & \text{when $k$ is even}, \\
-\frac{5}{8} + \frac{1}{2} \sum_i \left[ \nu_i(1-\nu_i) +\nut_i(1+\nut_i) \right] &
\text{when $k$ is odd}. \end{cases} \nonumber
\end{align}
The latter includes contributions from all the left-moving degrees of freedom --- the left-moving fermions $\xi$, the hidden $\GE_8$ (always in its NS sector), and the spacetime bosons in light-cone gauge.  Note that $q$ includes the contribution from $\gamma^6$.
To obtain the $\GUL'$ charge $q'$ we simply omit that contribution.

\subsection{Application to the M1 LG model}
In this case $\alpha_i = 1/5$, and  the charges of the twisted vacua $|k\ra$,
$k=0,\ldots,5$ are given in table~\ref{table:quincharge}.  The remaining
sectors contain CPT conjugate states and rarely require explicit
consideration.  The table lists both the $\GUL'$ and $\GUL$ charges,
where the computation of the former omits the $\gamma_6$ contribution to $q$ in~(\ref{eq:charges}).
\begin{table}
\begin{center}
\begin{tabular}{|c|c|c|c|c|c|c|}
\hline
$k$ 		&$E$	&$q'$		&$q$			&$\qb$ 		&$\nu$		&$\nut$			
\\ \hline
$0$		&$0$		&$-\ff{3}{2}$	&$-2$		&$-\ff{3}{2}$	&$0$			&$0$				
\\ \hline
$1$		&$-1$	&$0$			&$0$			&$-\ff{3}{2}$	&$\ff{1}{10}$	&$-\ff{2}{5}$		
\\ \hline
$2$		&$0$		&$\ff{3}{2}$	&$1$			&$-\ff{3}{2}$	&$\ff{1}{5}$	&$-\ff{4}{5}$		
\\ \hline
$3$		&$-\half$	&$-1$		&$-1$		&$-\ff{1}{2}$	&$\ff{3}{10}$	&$-\ff{1}{5}$		
\\ \hline
$4$		&$0$		&$\half$		&$0$			&$-\half$		&$\ff{2}{5}$	&$-\ff{3}{5}$		
\\ \hline
$5$		&$0$		&$-2$		&$-2$		&$\ff{1}{2}$	&$\ff{1}{2}$	&$0$				
\\ \hline
\end{tabular}
\caption{Vacuum quantum numbers for the quintic.}
\label{table:quincharge}
\end{center}
\end{table}

To compute the spectrum we apply the algorithm described in section \ref{ss:LGquintic}.
When $J_6 \in \la J_1,\ldots, J_5\ra$, we simply recover the familiar quintic spectrum.  So, without loss of generality
we may assume that $\cJ \equiv \la J_1,\ldots, J_5,J_6\ra$ is irreducible and zero-dimensional.\footnote{The
latter condition guarantees that we are off the singular locus.}  In what follows  we will assume that $\cJ$ is generic.

The classification of states is facilitated with some convenient notation.
We let $R = \C[\phi_1,\ldots,\phi_5]$, $\cF = R^{\oplus 6}$, and denote the degree $d$ components of the ring and module by $R_{[d]}$ and $\cF_{[d]}$.  We will only describe the most involved sectors $k=0$ and $k=1$, leaving
the remaining sectors for the amusement of the reader.

{\bf\noindent k=0:~~~}  The analysis in the untwisted sector reduces to the zero modes.  The supercharge is given
by
\begin{align}
\cQb = J_i (\phi) \gammab_i^\dag + J_6 (\phi) \gammab_6^\dag,
\end{align}
and the $E=0$ states are constructed from the $\phi_i$ and $\gammab_i,\gammab_6$ zero modes.
The states at fixed $q,\qb$ can be written as
\begin{align}
|\psi \ra = \left[ F_{[d]}^{i_1\cdots i_p } \gammab_{i_1}\cdots \gammab_{i_p}
                +G_{[d-1]}^{i_2 \cdots i_p} \gammab_{i_2} \cdots \gammab_{i_p} \gammab_6 \right]|0\ra,
\end{align}
and since we must project onto integral $q$,  $d = 5q-4p$.
We refer to this vector space as $\wedge^p \cF_{[d]}$, with $\wedge^0 \cF_{[d]} \equiv R_{[d]}$.  Its dimension
is given by
\begin{align}
\dim \wedge^p \cF_{[d]} = \binom{5}{p} \#(5,d) + \binom{5}{p-1} \#(5,d-1),
\end{align}
where $\#(n,d)$ is the dimension of $\Sym^d \R^n$,
\begin{align}
\#(n,d)=\begin{cases} \binom{n+d-1}{d} & d \ge 0 \\ 0 &\text{otherwise}. \end{cases}
\end{align}

 It is a general property
of (0,2) LG theories that the $\cQb$ cohomology in the untwisted sector is given by the homology of the Koszul
complex for the ideal $\cJ$~\cite{Kawai:1994qy}.  In the present case the
computation is considerably simplified when organized by degree.  The degree $d$ complex is
\begin{align}
\cK^{[d]}_{\bullet} = \xymatrix{\cdots \ar[r] & \wedge^{2} \cF_{[d-8]} \ar[r]^-{\cQb} & \cF_{[d-4]} \ar[r]^-{\cQb} &R_{[d]} \ar[r] & 0 ,}
\end{align}
with homology groups $H_{p}^{[d]}$.  Note that $H_{0}^{ [d]} = R_{[d]}/ \cJ$,
a simple generalization of the familiar LG chiral ring.
Since $\cJ$ is irreducible and zero-dimensional, the homology is concentrated
in the two right-most entries~\cite{Eisenbud:1995ca}.  We expect that the dimensions
of $H_{0}^{[d]}$ and $H_{1}^{[d]}$ are simply given by counting the dimensions of the $\wedge^p \cF_{[d]}$ vector
spaces.\footnote{It is not hard to convince oneself that the dimensions so obtained are correct for generic $\cJ$; for
instance, one can just adapt a similar argument from the large radius analysis of~\cite{Donagi:2006yf}.  Of course extra massless states do arise in less generic situations.}
Computing these dimensions, we find
\begin{align}
\dim H_0^{[5m]} &= \begin{cases}
1 & m=0   \leftrightarrow \rep{8}^c_{-2,-3/2} \\
100 & m=1  \leftrightarrow \rep{8}^{s\oplus 100}_{-1,-1/2} \\
0    & m > 1
\end{cases}, \quad
\dim H_1^{[5m]}  &= \begin{cases}
100 & m=3   \leftrightarrow \rep{8}^{s\oplus 100}_{1,1/2} \\
1    & m=4  \rightarrow \rep{8}^{c}_{2,3/2} \\
0    & \text{otherwise}
\end{cases}\label{eq:k0states_1}.
\end{align}
Clearly the two contributions are CPT conjugates of each other.

{\bf\noindent k=1:~~~}  We now repeat the exercise in the first twisted sector.  The main new feature
here is that we must identify states with internal energy $-1, -1/2,$ and $0$.
\begin{itemize}
\item[1.]  The unique $E=-1$ state $|1\ra$ can be used to construct the dilatino, gravitino, hidden $\GE_8$ gauginos,
as well as the $\SO(8)$ gauginos $\rep{28}_{0,-3/2}$.  The latter states are simply $\xi^a\xi^b|1\ra$.
\item[2.] The $E=-1/2$ states consist of  the gauginos $\gamma^6 |1\ra \leftrightarrow \rep{8}^v_{-1,-3/2}$,
as well as the $q=1$ states with $\qb = -3/2$ and $\qb=-1/2$:\footnote{We use the notation $|\psi\ra_{k}$ to
indicate the multiplicity $k$ of states of type $|\psi\ra$.}
\begin{align*}
\xymatrix{  \gammab_6|1\ra_1 \oplus \phi_i\gammab_j|1\ra_{25} \ar[r]^-{\cQb} & F_{[5]} |1\ra_{126}.}
\end{align*}
$\cQb$ has a trivial kernel, leading to $\rep{8}^{v\oplus 100}_{1,-1/2}$ states corresponding to
the $100$ elements of $R_{[5]}/\cJ$.
\item[3.] Finally, we examine the states with zero internal energy, which turn out to have $q=0$ or $q=2$.  The
$q=0$ states have $\qb=-3/2$ and $\qb=-1/2$:
\begin{align*}
\xymatrix{ {\begin{matrix} \gamma_6\gammab_6|1\ra_1&\oplus& \gamma_i\gammab_j|1\ra_{25} \\& \oplus& \\
 \phi_i\gammab_j\gamma_6|1\ra_{25}&\oplus&\phi_i\phib_j|1\ra_{25}\end{matrix}} \ar[rr]^-{\cQb} &&
 {F^i_{[4]} \gamma_i|1\ra_{350} \oplus \gamma_6 F_{[5]} |1\ra_{126}. }}
\end{align*}
Here $\cQb$ has a one-dimensional kernel, essentially due to the quasi-homogeneity of the LG superpotential.  Thus, we find a gaugino $\rep{1}_{0,-3/2}$ corresponding to the $\GUL$ symmetry, as well as
$401$ $\SO(10)$ singlets with $\qb=-1/2$.

The $q=2$ states are found at $\qb =-3/2,-1/2$ and $1/2$:
\begin{align*}
\xymatrix{ {\begin{matrix}  \phi_i\gammab_j\gammab_6|1\ra_{25} \\ \oplus \\ F^{ij}_{[2]} \gammab_i\gammab_j|1\ra_{150}\end{matrix}}
 \ar[rr]^-{\cQb} && {\begin{matrix} F^i_{[6]} \gammab^i |1\ra_{1050} \\ \oplus \\ F_{[5]} \gammab_6 |1\ra_{126} \end{matrix}}
 \ar[rr]^-{\cQb}   && {F_{[10]}|1\ra_{1001}. }}
\end{align*}
For generic $\cJ$ the cohomology is empty.
\end{itemize}
The analysis of the remaining sectors is, if anything, simpler, and combining all of the results leads to the
``puzzling'' spectrum described in section~\ref{ss:LGpuzzle}.

\subsection{Constraints from the permutation symmetry} \label{app:permsym}
In this section we will present a slight generalization of the
discrete R-symmetry arguments used in~\cite{Dine:1988kq}.
This generalization is based on the simple fact that
the LG presentation of the spectrum makes it easy to determine
the transformation properties of all the massless fields under
the discrete R-symmetries.\footnote{Although the reasoning involved
is fairly elementary, to our knowledge it has not been given explicitly.
Since it seems to be a fairly useful result in the context of LG vacua,
we will indulge in a little bit of detail.}  In this section we label states by
their $\SO(10)\times\GUL'$ representations.

Consider the LG theory on the five-dimensional $S_5$-preserving locus
of complex structures.  The $S_5$ is a global symmetry in the spacetime
theory, and we seek to determine
the action of permutations on the spectrum of massless fermions.
The massless fermions are presented as states $\cO |k\ra$, where $\cO$ is an operator
constructed from the first excited modes of the twisted fields, and the action
of a permutation $P$ takes the form
\begin{align}
P \cO |k\ra = e^{2\pi i \sigma_P(k)} \cO_P |k\ra,
\end{align}
where  $\cO_P$ is simply obtained by permuting the modes in $\cO$, and
$P|k\ra = e^{2\pi i\sigma_P(k)}|k\ra$.  Our task is to determine the $\sigma_P(k)$.

We begin with some intuition from the large radius analysis.  Under odd permutations
of the $\P^4$ coordinates the holomorphic $3$-form $\Omega$ changes sign, and
hence odd permutations should correspond to R-symmetries (see, for instance, section
16.5.3 of~\cite{Green:1987sp}).  Since $\Omega$, or rather, the corresponding chiral primary
operator in the SCFT, is the square of the right-moving spectral flow
generator $\Sigmab(\zb)$, we see that the odd permutation acts as
$P \Sigmab = \pm i \Sigmab P$.  Of course there is a similar action
on the left-moving spectral flow operator:
$P  \Sigma = \mp i \eta \Sigma P$, with $\eta =1 $ or $-1$.\footnote{
As we shall see shortly, the sign ambiguities will not affect our results.}
Our primary interest is in these R-symmetries, and we will restrict attention
to odd permutations henceforth.

The right-moving spectral flow operators show up in spacetime supercharges,
and in our conventions we have
\begin{align}
Q_\alpha =  \oint d\zb~ e^{-\vphi/2} \cS_\alpha \Sigmab^\dag (\zb),
\qquad Q_{\dot{\alpha}}  = \oint d\zb ~e^{-\vphi/2} \cS_{\dot{\alpha}} \Sigmab(\zb),
\end{align}
where $\vphi$ is the spin-field for the $\beta\gamma$ ghost system, while
the $\cS_{\alpha}$ and $\cS_{\dot{\alpha}}$ are the spin-fields for the
fermions in the (0,1) multiplets of the $\R^{1,3}$ degrees of freedom.  Thus,
$Q_\alpha$  ($Q_{\dot{\alpha}}$) carries charges $q=0$ and $\qb = -3/2$ ($\qb = 3/2$).
In particular, a gauge boson $v_\mu$ transforms under supersymmetry as
\begin{align}
\CO{\cQ_\alpha}{v^\mu} \sim \gamma^\mu_{\alpha \dot{\beta}} \lambdab^{\dot{\beta}},
\end{align}
and hence the gaugino $\lambdab$ (corresponding to a state with $\qb = -3/2$) should transform the same way as $Q_\alpha$ under $P$.  For example, the $\rep{45}$
gauginos represented by $\xi^\alpha\xi^\beta |1\ra$ in the LG spectrum\footnote{Recall that the $\xi^\alpha$ are the
ten free left-moving fermions.} should pick up a phase $\mp i$ under $P$, implying
$P  |1\ra  = \mp i |1\ra$.  Since the $\rep{16}_{-3/2}$
and $\brep{16}_{3/2}$ gauginos are related to the gauginos in the $k=1$ sector by left-moving
spectral flow, we also find
\begin{align}
P  |0\ra =  \eta |0\ra, \qquad P  |2 \ra = -\eta |2\ra.
\end{align}
Although $P$ does not act homogeneously on the $\GE_6$ representations, it is easy to
combine it with a $\GUL'$ gauge transformation to define an action $\Pt = P e^{\mp i \eta \pi J'}$
that does act homogeneously on the gauginos,
$\Pt : \lambdab \mapsto \mp i \lambdab$.
This is clearly an R-symmetry, with the spacetime superspace coordinates transforming as
$\theta^\alpha \mapsto \pm i \theta^\alpha$.  Note that $\CO{\Pt}{\Sigma} = 0$.

Next, we consider the chiral (i.e. $\qb =-1/2$) $\brep{27}$ states.  At the Gepner point the $101$
$\rep{10}_1$  in the $k=1$ sector have the form $\xi^\alpha \cO |1\ra$, where $\cO$ is a
quintic polynomial in the $\phi_i$.  Hence, $\Pt : \rep{10}_1 \mapsto \pm i P_0(\rep{10}_1)$,
where $P_0$ simply permutes the $\phi_i$.  The remaining components of
$\brep{27}$ are related by left-spectral flow, and it is not hard to verify that
$\Pt :\brep{27} \mapsto \pm i P_0 (\brep{27})$.

The transformation of the single $\rep{10}_{-1}$ is completely determined by that of $|3\ra$.
The $k=3$ sector also contains the superpartner of the K\"ahler modulus, represented by
$\sum_i \phi_i\gammab_i |3\ra$.  Since
the K\"ahler modulus is invariant under all permutations, its chiral superpartner must transform
oppositely to $\theta^\alpha$, which implies $\Pt |3\ra =\pm i |3\ra$.  Since left-moving spectral
flow relates $|3\ra$ to $|4\ra$ and $|4\ra$ to $\gammab^1\cdots \gammab^5 |5\ra$, we see
that $\Pt : \rep{27} \mapsto \pm i \rep{27}$.  The $\Pt$ transformations of
$|1\ra$ and $|3 \ra$ imply that all of the $\GE_6$ singlets transform as
$\Pt: S \mapsto \mp i P_0(S)$.

We have now determined the transformations of all the chiral fermions as well as $\theta^\alpha$,
and from this we easily obtain the transformations of the chiral superfields under odd permutations:
\begin{align}
\Pt : \Phi_{\rep{27}} \mapsto - \Phi_{\rep{27}}, \qquad
\Pt :\Phi_{\brep{27}}\mapsto - P_0 (\Phi_{\brep{27}}), \qquad
\Pt : \Phi_S \mapsto P_0 (\Phi_S).
\end{align}
Of course the superpotential must transform as $\theta^2$, i.e. $W \mapsto - W$.  But now we simply
observe that the terms $(\Phi_{\rep{27}} \cdot \Phi_{\brep{27}^0})^k$ and $\Phi_S (\Phi_{\rep{27}} \cdot \Phi_{\brep{27}^0})^k$,
where $\brep{27}^0$ corresponds to the unique permutation-invariant monomial $x_1\cdots x_5$, cannot appear in $W$.  Thus, for any point on the $S_5$ locus we have found a class of unobstructed
deformations Higgsing $\GE_6 \to \SO(10)$.  Since the symmetry and the charge assignments persist for any
value of the K\"ahler modulus, we can extend this to large radius as well.  That is of course what we might expect
from the supergravity analysis of~\cite{Li:2004hx}.

Let us finally note that the phases $e^{2\pi i\sigma_P(k)}$ can be determined without appealing to any
large radius intuition.\footnote{We thank S.~Hellerman for discussions about this point.}  Consider,
for instance, the permutation exchanging the chiral superfields $X_1$ and $X_2$ in some (2,2) LG orbifold.  In order for this
to be a symmetry, the two fields must have equal weight, and therefore also $\nu_1 = \nu_2$ and
$\nut_1 = \nut_2$.  We can then think of the permutation as a global $\Z_2$ symmetry that commutes
with the Landau-Ginzburg orbifold and leaves $X_1+X_2$ invariant, while changing the sign of $X_1-X_2$.
As discussed in~\cite{Aspinwall:2010ve}, the phase of the twisted vacuum $|k\ra$ under such a global
symmetry is given by
\begin{align}
\sigma_P(k) = \ff{1}{2} (\nut_1 -\nu_1+1) \pmod 1.
\end{align}
Applying this to the quintic, it is easy to check that this leads to the same results as above.

\section{The low energy, $r\gg 0$ limit of the quintic and M2 GLSMs}
\label{app:m1red}
In this appendix we will try to recover the (2,2) locus in the NLSM derived from the
M2 GLSM in the classical low energy limit.  To obtain this limit, we must take the dimensionful couplings of the gauge theory to infinity.  These naturally include the gauge coupling $e$, as well as the superpotential couplings.  The resulting low energy theory receives quantum corrections that are suppressed in the limit $r \gg 0$.  We will compare the low energy actions obtained from the quintic and M2 GLSMs.

In the $e \to \infty$ limit the fluctuations of the gauge multiplets are suppressed, while the D-term is imposed as a constraint on the fields:  the light bosons must satisfy
\begin{align}
 \phi_i\phib_i - 5 \phi_0 \phib_0 - r =0.
\end{align}
 In addition, when $P$ is a non-singular hypersurface (i.e. $P_{,i} = 0$ has no solutions in $\P^4$), the bosons are further constrained to $\phi_0 = 0$ and $P(\phi) = 0$.  To solve these constraints, we work in a patch with $\phi_1 \neq 0$ and define affine coordinates $Z^I = \phi_I / \phi_1$, $I=2,\ldots,5$.  Fixing the gauge to $\arg \phi_1 = 0$, the D-term constraint is solved by
 \begin{align}
 \phi_1 = \sqrt{\frac{r}{1+Z\cdot\Zb}}~,   \qquad
 \text{with} \quad
 Z \cdot \Zb \equiv Z^I \Zb^I.
 \end{align}
Since we must also demand $P(Z) = 0$, we choose a parametrization $Z^I(z^a)$, $a=1,2,3$ for solutions to $P(Z) = 0$ in the $\phi_1\neq 0$ patch.

In the quintic GLSM the fluctuations of $\sigma$ are also suppressed in the $e\to \infty$ limit, and we
can eliminate it via its algebraic equation of motion:  $\sigma = i \psib^i\gamma^i / r\sqrt{2}$.  The remaining terms in the action are then a sum of
\begin{align}
\label{eq:quinticaction}
\cL_{\text{kin}} &= \ff{1}{2}  \nabla_+\phib_i \nabla_-\phi_i
+ \ff{1}{2} \nabla_-\phib_i \nabla_+\phi_i +i \psib^i  \nabla_- \psi^i + i\gammab^i  \nabla_+\gamma^i
\nonumber\\
&~~+i \psib^0  \nabla_- \psi^0 + i\gammab^0  \nabla_+\gamma^0   ,\\
\cL_{\text{Yuk}} &=  \lambdab_-\psib^i\phi_i
+  \alpha \lambda_+ \gammab^i\phi_i
+\gamma^0 P_{,i} \psi^i + \gamma^iP_{,i} \psi^0
+\text{h.c.}\nonumber\\
\cL_{4} &= \alpha r^{-1} \gammab^i\psi^i \psib^i\gamma^i.\nonumber
\end{align}
The parameter $\alpha$ is introduced to distinguish the quintic ($\alpha =1$) and M2 ($\alpha = 0$) GLSMs.

The bosonic action of the NLSM is obtained by integrating out the $v_\pm$ gauge field.  This leads to
\begin{align}
\cL^B = \p_+\phi_1\p_-\phi_1 + \p_+\phi_I \p_-\phib_I -r^{-1} J^B_+J^B_-,
\end{align}
where $J^B_\pm$ are the bosonic terms in the $\GU(1)$ gauge current
\begin{align}
J^B_\pm = \ff{i}{2} \phi_1^2 (Z\cdot \p_\pm \Zb - \Zb \cdot \p_\pm Z).
\end{align}
A slightly tedious computation yields the expected result:
\begin{align}
\cL^B = \ff{1}{2} h_{\ab b}  (\p_+ \zb^{\ab} \p_-z^{b} +\p_- \zb^{\ab} \p_+ z^b ),
\end{align}
where $h_{\ab b}$ is the pull-back of the $\P^4$ Fubini-Study metric to $M$ in the $\phi_1\neq 0$ patch.

The fermionic action is a bit more interesting.  To simplify its form, we introduce a change of basis
\begin{align}
\label{eq:Fbasis}
\gamma^i = \phi_i \chi + \frac{\Pb_{,i}}{ |\Pb_{,j}|} \chi' +
S^i_a \chi^a, \qquad
\psi^i  =  \phi_i \eta + \frac{\Pb_{,i}}{ |\Pb_{,j}|} \eta' +
S^i_a \eta^a,
\end{align}
where the $S^i_a(z,\zb)$ are given by
\begin{align}
S^1_a = -S^I_\alpha \Zb^I, \qquad
S^I_a = \phi_1
\left( \delta^I_J - \frac{Z^I \Zb^J}{1+Z\cdot \Zb}\right)
\frac{\p Z^J}{\p z^a},
\end{align}
and satisfy
\begin{align}
\label{eq:sreq}
\phib_i S^i_a = 0, \qquad
P_{,i} S^i_a = 0, \qquad
\Sb^i_{\ab} S^i_b = h_{\ab b}.
\end{align}
With this choice of basis the Yukawa terms in~(\ref{eq:quinticaction}) simplify to
\begin{align}
\cL_{\text{Yuk} }= r \lambdab_- \etab
+r\alpha \lambda_+ \chib
+|P_{,i}| (\gamma^0 \eta' +\chi' \psi^0) + \text{h.c.}~.
\end{align}
We now see that $\lambdab_-$ is a Lagrange multiplier for
$\etab = 0$, and similarly, when $\alpha\neq0$,  integration over
$\lambda_+$ will force $\chib = 0$.  Since $P$ is non-singular, in the low energy limit the last term gives large masses to the $\gamma^0$, $\eta'$, $\chi'$ and $\psi^0$ fermions.  We can therefore set these excitations to zero.  Remembering to include the fermions' contributions to the equations of motion for $v_\pm$ gauge fields, we find
\begin{align}
\label{eq:LF}
\cL^F &= i h_{\ab b} \left[ \etab^{\ab} D_- \eta^b + \chib^{\ab} D_+\chi^b \right] + i r\chib\p_+\chi + i (\chib^{\ab} \Sb^i_{\ab} \p_+\phi_i \chi + \chib \phib_i \p_+ S^i_a \chi^a) \cr
~ &~~-r^{-1} h_{\ab b} \etab^{\ab} \eta^b
(r \chib \chi +h_{\cb d} \chib^{\cb} \chi^d)
+\alpha r^{-1} h_{\ab b} h_{\cb d} \chib^{\ab}\eta^{b} \etab^{\cb}\chi^{d}
+\alpha (\lambda_+\chib + \lambdab_+\chi),
\end{align}
where
\begin{align}
\label{eq:Tconnection}
D_-\eta^b &= \p_- \eta^b + h^{b\bb} \Sb^i_{\bb} \p_- S^i_c \eta^c -i r^{-1} J^B_- \eta^b = \p_-\eta^b + \Gamma^b_{cd} \p_- z^c \eta^d
\nonumber\\
D_+\chi^b &= \p_+ \chi^b + h^{b\bb} \Sb^i_{\bb} \p_+ S^i_c \chi^c -i r^{-1} J^B_+ \chi^b = \p_+\chi^b +\Gamma^b_{cd} \p_+ z^c \chi^d
\end{align}
are the usual covariant derivatives with the Christoffel connection for the metric $h_{\ab b}$.

Setting $\alpha = 1$, we obtain the fermion action for the quintic GLSM:
\begin{align}
\cL^F = i h_{\ab b} (\chib^{\ab} D_+\chi^b + \etab^{\ab} D_- \eta^b)
-r^{-1} (h_{\ab d} h_{\cb b} +h_{\ab b} h_{\cb d}) \etab^{\ab} \eta^b
\chib^{\cb} \chi^d.
\end{align}
The four-fermi term is the Riemann tensor for $h_{\ab b}$, and $\cL^B+\cL^F$ is just the familiar (2,2) NLSM action for the quintic with induced metric $h_{\ab b}$.

Setting $\alpha = 0$, we obtain the M2 model for $J_i = P_{,i}$.  As far as the bosons and right-moving fermions are concerned, we of course obtain the same form as the quintic action; however, the left-moving degrees of freedom are markedly different, as this is genuinely a (0,2) NLSM.  To describe the action, we let $\alpha,\beta = 0,1,2,3$ and define the following diagonal metric on $\cO_M \oplus T_M$
\begin{align}
H_{\alphab \beta} = \begin{pmatrix} r & 0 \\ 0 & h_{\ab b} \end{pmatrix},
\end{align}
as well as connection $\cA = \cA_{a} dz^a + \cA_{\ab} d\zb^{\ab}$,
with
\begin{align}
\cA^\beta_{a \gamma} = \begin{pmatrix}
0 & 0 \\
\delta^b_a & \Gamma^b_{ac} \end{pmatrix}, \qquad
\cA^\beta_{\ab \gamma} = \begin{pmatrix}
0 & -r^{-1} h_{\ab c} \\
0 & 0 \end{pmatrix}.
\end{align}
This connection has a $(1,1)$ curvature two-form $\cF$,
\begin{align}
(\cF_{\ab b})_{\gammab \delta} = H_{\gammab\gamma} (\cF_{\ab b})^\gamma_\delta
= r^{-1} h_{\ab b} \begin{pmatrix}
r & 0 \\ 0 & h_{\cb d} \end{pmatrix},
\end{align}
and the left-moving part of the action is written as
\begin{align}
\cL^{\chi}_{\text{M2}} = i H_{\alphab \beta} \chib^{\alphab} D'_+ \chi^\beta + \etab^{\ab} \eta^b (\cF_{\ab b})_{\gammab \delta} \chib^{\gammab} \chi^\delta,
\end{align}
where $D'_+$ is defined with the pull-back of the connection $\cA$.

Clearly the M2 GLSM with $J_i = P_{,i}$ does not reduce to the naive expectation of a (2,2) quintic NLSM and a free left-moving fermion $\chi$.  Since $\cF$ has full rank, we cannot obtain the desired theory by a field redefinition.  Of course there is an easy way to obtain the expected theory by varying $\cA$.
We write
\begin{align}
\cA = \cAh + t \cB,  \qquad
\cB_{a} = \begin{pmatrix}  0 & 0 \\ \delta^b_a & 0 \end{pmatrix},\qquad
\cB_{\ab} = -r^{-1}\begin{pmatrix} 0 & h_{\ab c} \\ 0 & 0 \end{pmatrix},
\end{align}
so that the (2,2) quintic NLSM (with a free left-moving fermion) is obtained at $t=0$, while the M2 NLSM is found at $t=1$.

It is instructive to linearize the deformation of the action at $t=0$.
The quintic NLSM is invariant under a (0,2) supersymmetry, with non-zero $\cQb$ variations
\begin{align}
\cQb\cdot \zb^{\ab} = \etab^{\ab}, \qquad
\cQb\cdot \eta^a = i  \p_+ z^a, \qquad
\cQb \cdot \chib^{\ab} = - \Gamma^{\ab}_{\bb \cb} \etab^{\bb} \chib^{\cb},
\end{align}
Linearizing $\cL^\chi_{\text{M2}}$ around $t=0$, we find $\delta_t \cF = 0$, and $\delta_t \cL^F = t \cO_{\cB}$, with
\begin{align}
\cO_{\cB} = i h_{\ab b} (\chib^{\ab} \p_+ z^b \chi -\chib \p_+\zb^{\ab} \chi^b) = \cQb \cdot ( - h_{\ab b} \chib^{\ab} \eta^b \chi) + \text{h.c.}~.
\end{align}
It is not hard to check that the resulting $\delta_t S$ is $\cQb$-closed up to equations of motion of the left-moving fermions.
This deformation is not $\cQb$-exact, and in fact takes the form of a (0,2) superpotential coupling.
Turning on the $\cO_{\cB}$ coupling has a clear geometric
significance.  Since $\delta_t \cF = 0$, and the change in $\cA$ is not pure gauge, this is a deformation of $\cO\oplus T_M$, and a look at $\cB$ will convince the reader that this is a deformation by an element of $H^1(T^\ast_M)$ without an accompanying $H^1(T_M)$ element.  Such a deformation cannot by itself lead to a stable bundle, and
thus, the reduction of the M2 model with $J_i = P_{,i}$ naively appears to be destabilized.

We do not think this means that the large radius (2,2) locus cannot be recovered
in the context of the M2 model; however, our simple analysis certainly suggests that
the limit $J_i \to P_{,i}$ has some subtleties that remain to be understood properly.
For instance, it might be that  (2,2) supersymmetry only emerges as an accidental
IR symmetry.
Of course none of these issues arise in the M1 model.  There we can identify the (2,2) locus in terms of GLSM parameters, and this simple clarification is certainly worth the price of extra fields in the UV.

\end{appendix}

\bibliographystyle{my-h-elsevier}

\end{document}